\documentclass[12pt, draftclsnofoot, onecolumn]{IEEEtran}

\usepackage{amsmath}
\usepackage{amssymb}
\usepackage{graphicx}
\usepackage{float}
\usepackage{fixltx2e}
\usepackage{subeqnarray}
\usepackage{cases}
\allowdisplaybreaks
\IEEEoverridecommandlockouts
\usepackage[square, comma, sort&compress, numbers]{natbib}
\usepackage{multirow}
\usepackage{stfloats}
\usepackage{fixltx2e}
\usepackage[font={small}]{caption}
\usepackage{epsfig}
\usepackage{epstopdf}
\usepackage{optidef}

\newtheorem{Theorem}{Theorem}

\usepackage{algorithm}
\usepackage{algpseudocode}
\ifCLASSINFOpdf
\else
\fi
\usepackage{amsmath}
\usepackage{array}
\usepackage{fixltx2e}

\usepackage{stfloats}

\begin{document}

\title{Cooperative Asynchronous Non-Orthogonal Multiple Access with Power Minimization Under QoS Constraints\\}

\author{\IEEEauthorblockN{Xun Zou, \textit{Student Member, IEEE}, Mehdi~Ganji, \textit{Student Member, IEEE}, and Hamid~Jafarkhani, \textit{Fellow, IEEE}}\\
	\thanks{This work was supported in part by the NSF Award CCF-1526780. The authors are with the Center for Pervasive Communications and Computing, Department of Electrical Engineering and Computer Science, University of California, Irvine, CA, 92697 USA (email: \{xzou4, mganji, hamidj\}@uci.edu).}
}

\maketitle
\begin{abstract}
Recent studies have demonstrated the superiority of non-orthogonal multiple access (NOMA) over orthogonal multiple access (OMA) in cooperative communication networks. In this paper, we propose a novel half-duplex cooperative asynchronous NOMA (C-ANOMA) framework with user relaying, where a timing mismatch is intentionally added in the broadcast signal. We derive the expressions for the individual throughputs of the strong user (acts as relay) which employs the block-wise successive interference cancellation (SIC) and the weak user which combines the symbol-asynchronous signal with the interference-free signal. We analytically prove that in the C-ANOMA systems with a sufficiently large frame length, the strong user attains the same throughput to decode its own message while both users can achieve a higher throughput to decode the weak user's message compared with those in the cooperative NOMA (C-NOMA) systems. Besides, we obtain the optimal timing mismatch when the frame length goes to infinity. Furthermore, to exploit the trade-off between the power consumption of base station and that of the relay user, we solve a weighted sum power minimization problem under quality of services (QoS) constraints. Numerical results show that the C-ANOMA system can consume less power compared with the C-NOMA system to satisfy the same QoS requirements.
\end{abstract}


\section{Introduction}
Non-orthogonal multiple access (NOMA) has been regarded as one of the key technologies for the next generation wireless communications~\cite{3gpp2018noma}. Compared with the conventional orthogonal multiple access (OMA), NOMA can provide massive connectivity and high spectral efficiency~\cite{ding2017survey}. The key rationale behind NOMA is to allow users to share non-orthogonal wireless resources, e.g., frequency, time, and code. For multiuser detection, the superposition coding and the successive interference cancellation (SIC) are employed at the transmitter and receiver, respectively. 

Cooperative communication is an effective approach to exploit spatial diversity available through cooperating terminals' relaying signals for one another~\cite{laneman2004cooperative,koyuncu2012distributed,jing2009single}. Cooperative relaying network with NOMA has been extensively studied in the literature, e.g.,~\cite{kim2015capacity,men2015non,liang2017outage}. It has been shown that the cooperative NOMA (C-NOMA) systems outperform the cooperative OMA systems in terms of the spectral efficiency~\cite{kim2015capacity} and the outage probability~\cite{men2015non}. Instead of dedicated relay nodes, users can also be adopted as relays in a cooperative network. A key feature of NOMA is that users with better channel conditions have prior information about the messages of other users. Ding et al.~\cite{ding2015cooperative} proposed a C-NOMA scheme to fully exploit the prior knowledge at the strong user, where the users could cooperate with each other via short-range communication channels. Yue et al.~\cite{yue2018exploiting} compared different operation modes of the relay user in a C-NOMA system. The half-duplex relay user receives and transmits in separate time slots while the full-duplex relay user receives and transmits simultaneously. In~\cite{yue2018exploiting}, the outage probability, the ergodic rate, and the energy efficiency were analyzed in a NOMA user relaying system where the near user could switch between full-duplex and half-duplex modes to relay messages to the far user. Zhang et al.~\cite{zhang2017full} studied an adaptive multiple access scheme to further improve the outage performance, which dynamically switched among the C-NOMA with user relaying, conventional NOMA, and OMA schemes, according to the level of residual self-interference and the quality of links. Wei et al.~\cite{wei2018energy} solved the energy efficiency maximization problem of a full-duplex C-NOMA system under the constraint of successful SIC operation. 

\subsection{Motivations and Related Works}

By intentionally introducing symbol asynchrony in the transmitted signal, asynchronous NOMA (ANOMA) systems can achieve a better throughput performance compared with the conventional (synchronous) NOMA systems~\cite{zou2018analysis,cui2017asynchronous,ganji2019time}. In ANOMA systems, the receiver utilizes the oversampling technique~\cite{poorkasmaei2015asynchronous} to achieve the sampling diversity gain. It has been revealed that the cooperative communication systems can also benefit from the symbol-asynchronous transmission. Sodagari et al.~\cite{sodagari2018enhanced} studied an asynchronous cognitive radio framework, where the primary user and the secondary user were not aligned in their timing. They conclude that not only can asynchronous cognitive radio reduce the interference to the primary user, but it also saves power at the secondary user compared with synchronous cognitive radio systems. An asynchronous network coding (ANC) transmission strategy for multiuser cooperative networks was investigated in \cite{zhang2017exploiting,zhang2017asynchronous}, where the received signals from multiple sources were asynchronous to each other. The proposed scheme achieves full diversity and outperforms the complex field network coding in terms of decoding complexity and bit error rate (BER).

In this paper, we consider a half-duplex cooperative ANOMA (C-ANOMA) system with user relaying, including a base station (BS), a strong user (also acting as a relay), and a weak user. 
Different from the conventional C-NOMA systems, a symbol asynchrony is intentionally added to the downlink superposed signal in the broadcast phase of C-ANOMA systems. The weak user receives two blocks of signals via the broadcast link and the relay link separately. The questions then arise: How to realize SIC based on the symbol-asynchronous signal and then evaluate the performance of the strong user in the C-ANOMA systems? How to evaluate the performance of the weak user which combines a symbol-asynchronous signal from the broadcast link with an interference-free signal from the relay link? Moreover, compared with the cooperative systems with dedicated relay nodes, the power control strategy plays a more critical role in the cooperative systems with user relaying because the power consumption of the relay user affects the lifetime of the cooperative network. We assume that the channel information is available at transmitters~\cite{wei2018energy,liu2018hybrid} and the system works in the delay-tolerant transmission mode~\cite{yue2018exploiting}, such that the transmitters can dynamically adjust their transmit powers according to the channel states to avoid outage and save energy. On the one hand, the relay user with very limited battery capacity is more sensitive to the power consumption compared with BS. On the other hand, the relay user can transmit signals to the weak user more efficiently because the relay user is usually closer to the weak user. As a result, an effective power control strategy is of practical interest to make a trade-off between the transmit power of BS and that of the relay user while satisfying the quality of service (QoS) constraints in the C-ANOMA/C-NOMA systems with user relaying. To reduce the energy consumption, the power minimization problem has been investigated in several systems, e.g., the downlink NOMA systems~\cite{lei2016power}, the multicell NOMA systems~\cite{fu2017distributed}, and the cooperative beamforming networks~\cite{jing2009network}. Besides, Liu et al.~\cite{liu2018hybrid} and Chen et al.~\cite{chen2017power} studied the power allocation problem for half-duplex and full-duplex C-NOMA systems, respectively, to maximize the minimum achievable user rate in a NOMA user pair. To the best of our knowledge, the power minimization problem has never been studied in the C-NOMA or C-ANOMA systems with user relaying.

\subsection{Contributions}
In this paper, we comprehensively investigate a half-duplex C-ANOMA system with user relaying. The primary contributions of the paper are summarized as follows:
\begin{itemize}
\item We introduce the block-wise SIC technique into C-ANOMA systems, which is employed at the strong (relay) user. We derive the analytical expressions for throughputs achieved by the strong user to decode both users' messages and study their asymptotic performances as the frame length goes to infinity. We analytically show that in the C-ANOMA systems with a sufficiently large frame length, the strong user can achieve a higher throughput to detect the weak user's message while attains the same throughput when detecting its own message compared with those in C-NOMA systems.

\item We derive the expression for the throughput achieved by the weak user which combines the asynchronously superimposed signal from the broadcast link with the interference-free signal from the relay link. Based on the derived throughput expressions, we obtain the asymptotic throughput as the frame length goes to infinity and its simple upper and lower bounds. We analytically prove that in the C-ANOMA systems with a sufficiently large frame length, the throughput of the weak user is greater than that in the C-NOMA systems.

\item We further study the optimal design of C-ANOMA systems. We analytically prove that the optimal timing mismatch to maximize the individual throughput converges to half of the symbol interval as the frame length increases. Besides, we solve the weighted sum power minimization problem under the QoS constraints for C-ANOMA and C-NOMA systems. The solution is given by the explicit expressions of the powers allocated to the strong and weak users at BS and the transmit power of the relay (strong) user. It is demonstrated that for a relatively large frame length, the C-ANOMA systems consume less power compared with the C-NOMA systems in order to satisfy the same QoS requirements. In other words, under the same transmit power limits, the C-ANOMA systems can provide a higher QoS for users compared with the C-NOMA systems.
\end{itemize}

\subsection{Organization and Notation}
The remainder of the paper is organized as follows. The C-ANOMA system model is presented in Section \ref{sectionSysModel}.
The throughput performance of the C-ANOMA system is analyzed in Section \ref{secPerformanceAnalysis}. We discuss the optimal design of the C-ANOMA system in Section \ref{sectionSysdesign} where we investigate the optimal timing mismatch and solve the weighted power minimization problem under QoS constraints. Numerical results are presented in Section \ref{sec:numericalResults}. Finally, we draw the conclusions in Section \ref{sectionConclusion}.

Notations:
$(\cdot)^H$ denotes the Hermitian transpose, $(\cdot)^T$ denotes the transpose, $(\cdot)^{-1}$ denotes the inverse operation, $\otimes$ denotes the Kronecker product,
$|x|$ denotes the absolute value of $x$, $\bar{x}$ denotes the complex conjugate of $x$,
$\mathbb{E}[\cdot]$ denotes the expectation operation,
$\mathcal{CN}\left(0,1\right)$ denotes the complex normal distribution with zero mean and unit variance. $\mathrm{diag}(\mathbf{x})$ stands for a diagonal matrix whose $k$-th diagonal element is equal to the $k$-th entry of vector $\mathbf{x}$.

\section{System Model}\label{sectionSysModel}
\begin{figure}[t b]
	\centering
	\includegraphics[width=3in]{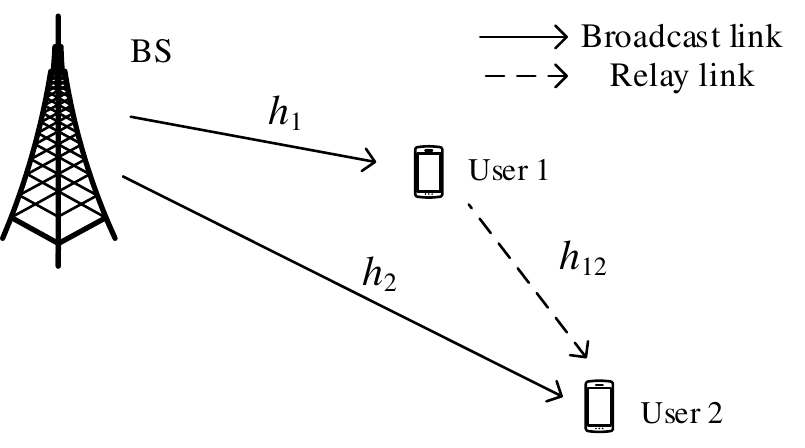}
	\caption{Illustration of a three-node C-ANOMA/C-NOMA system with user relaying.}
	\label{downlink}
\end{figure}
\begin{figure}[t b]
	\centering
	\includegraphics[width=4in]{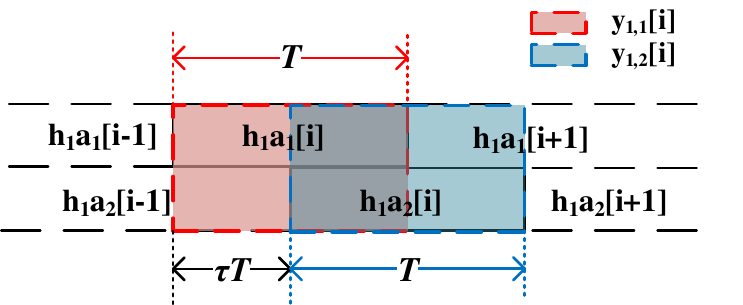}
	\caption{Illustration of the sampling for the broadcast phase in C-ANOMA systems.}
	\label{samplingnoerror}
\end{figure}

In this paper, as shown in Fig. \ref{downlink}, we consider a downlink half-duplex C-ANOMA system which includes a single BS and two users which are equipped with a single antenna. User 1 (strong user) acts as a relay for User 2 (weak user) and adopts the decode-and-forward (DF) protocol, i.e., decodes and forwards the message to User 2 via the relay link. The downlink transmission is done in blocks, including two phases, i.e., the broadcast phase and the relay phase. In the broadcast phase, BS broadcasts one block of superposed signal to two users simultaneously while User 1 is silent. In the relay phase, User 1 transmits the block of decoded signal to User 2 while BS keeps silent. We assume that the channel is static within each block~\cite{liu2016cooperative} and all the channel information is perfectly known at BS, Users 1 and 2~\cite{liu2018hybrid,wei2018energy}. The channel coefficient between BS and User $i$ is denoted as $h_i$ ($i=1, 2$) and the channel coefficient between Users 1 and 2 is denoted as $h_{12}$. In what follows, we present our analysis in the broadcast phase and the relay phase separately.

\subsection{Broadcast Phase}
\subsubsection{C-ANOMA}
In the C-ANOMA systems, a symbol mismatch is intentionally introduced in the downlink signal. As shown in Fig. \ref{samplingnoerror}, the intended timing mismatch between the symbols for Users 1 and 2 is denoted by $\tau T$, where $T$ is the symbol interval and $\tau$, $0 \le \tau < 1$, is the normalized timing mismatch. We assume that $\tau$ can be perfectly known at users as the downlink control information. Note that the C-ANOMA system becomes a synchronous C-NOMA system when $\tau = 0$. 

Let $a_1[i] = \sqrt{P_1}s_1[i]$ and $a_2[i]=\sqrt{P_2}s_2[i]$,
where $s_j[i]$ denotes the $i$th symbol sent to User $j$, $j = 1, 2$, $P_j$ stands for the power allocated to User $j$. The transmitted signal at BS is given by
\begin{align}
s(t) = \sum_{i = 1}^{N} a_1[i]p(t - iT) + \sum_{i = 1}^{N}a_2[i]p(t - iT - \tau T).
\end{align}
where $N$ denotes the number of symbols in a frame, i.e., the frame length, $p(\cdot)$ denotes the pulse-shaping filter. Without loss of generality, the rectangular pulse shape is adopted, i.e., $p(t)=1/\sqrt{T}$ when $t \in [0, T]$ and $p(t) = 0$ otherwise. 

The received signal at User 1 is given by
\begin{equation}\label{}
y_1(t) = h_1s(t) + n(t) =  h_1\left(\sum_{i = 1}^{N} a_1[i]p(t - iT) + \sum_{i = 1}^{N}a_2[i]p(t - iT - \tau T)\right) + n_1(t),
\end{equation}
where $n_1(t) \sim \mathcal{CN}(0, 1)$ denotes the normalized additive white Gaussian noise (AWGN). 

The oversampling technique~\cite{xunGlobecom16,xunICC18,ganji2016interference}, depicted in Fig. \ref{samplingnoerror}, is employed at the receiver to take advantage of sampling diversity in asynchronous systems. As shown in Fig. \ref{samplingnoerror}, the receiver uses the matched filter, sampling at $iT$ and $(i + \tau)T$, $i=1, \cdots, N$, to obtain two sample vectors, denoted by $[y_{1,1}[1], \cdots, y_{1,1}[N]]^T$ and $[y_{1,2}[1], \cdots, y_{1,2}[N]]^T$. Specifically, the $i$th element in the first sample vector is given by
\begin{align}\label{y_1i_noerror}
y_{1,1}[i] =& \int_{0}^{\infty}y_1(t)p(t - iT ) dt\notag\\
=& \int_{0}^{\infty}h_1a_1[i] p(t-iT)p(t - iT ) dt\notag\\
& + \int_{0}^{\infty} \left\{h_1a_2[i-1] p(t-(i+1+\tau) T) + h_1a_2[i] p(t-(i+\tau) T)\right\}p(t - iT ) dt + n_{1,1}[i] \notag\\
=& h_1a_1[i] + \tau h_1a_2[i-1] + (1 - \tau) h_1 a_2[i] + n_{1,1}[i],
\end{align}
where $n_{1,1}[i]= \int_{0}^{\infty}n_1(t) p(t-iT)dt$ denotes the additive noise. The $i$th element in the second sample vector is given by
\begin{align}\label{y_2i_noerror}
y_{1,2}[i] &= \int_{0}^{\infty}y_1(t)p(t - iT -\tau T) dt = h_1a_2[i] + \tau h_1a_1[i+1] + (1 - \tau) h_1a_1[i] + n_{1,2}[i],
\end{align}
where $n_{1,2}[i]=\int_{0}^{\infty}n_1(t) p(t-iT - \tau T)dt$ denotes the additive noise.

We can write the outputs of the two matched filters at User 1 in a matrix form as
\begin{align}\label{eqU1receivedsampleX1+X2}
\mathbf{Y}_1 &= h_1\sqrt{P_1}\mathbf{RG_1S_1} + h_1\sqrt{P_2}\mathbf{RG_2S_2} + \mathbf{N}_1
\end{align}
where $\mathbf{Y}_1 = \left[y_{1,1}[1]\ y_{1,2}[1]\ \cdots\ y_{1,1}[N]\ y_{1,2}[N]\right]^T$, $\mathbf{G}_1$ and $\mathbf{G}_2$ are $2N$-by-$N$ matrices given by $\mathbf{G_1} = \mathbf{I}_N \otimes [1\ 0]^T$ and $\mathbf{G_2}  = \mathbf{I}_N \otimes [0\ 1]^T$, $\mathbf{S_i} = \left[s_i[1]\  \cdots\ s_i[N]\right]^T$ ($i= 1, 2$), $\mathbf{N}_1 = \left[n_{1,1}[1]\ n_{1,2}[1]\ \cdots\ n_{1,1}[N]\ n_{1,2}[N]\right]^T$, and
\begin{align}
\mathbf{R} &= \left[\begin{smallmatrix}
1\ &1-\tau\ &0\ &\cdots\ &\cdots\ &0\\
1-\tau\ &1\ &\tau\ &0\ &\cdots\ &0\\
0\ &\tau\ &1\ &1-\tau\ &\cdots\ &0\\
\vdots\ &\ddots\ &\ddots\ &\ddots\ &\ddots\ &\vdots\\
0\ &\cdots\ &0\ &\tau\ &1\ &1-\tau\\
0\ &\cdots\ &\cdots\ &0\ &1-\tau\ &1
\end{smallmatrix}\right].\label{R_matrix}
\end{align}
Note that multiplying $\mathbf{R}$ by $\mathbf{G_i}$ outputs a $2N$-by-$N$ matrix whose columns are equal to the odd (if $i = 1$) or even (if $i = 2$) columns of $\mathbf{R}$.

We assume that the transmitted symbols are normalized and independent to each other, such that $\mathbb{E}\left[\mathbf{SS}^H\right] = \mathbf{I}$. Note that the noise terms in \eqref{y_1i_noerror} and \eqref{y_2i_noerror} are colored due to the oversampling, and we have
\begin{align}
\mathbb{E}\left\{n_1[i]n^H_2[i]\right\} =
\int_{0}^{\infty}\int_{0}^{\infty}
\mathbb{E}\left\{n_1(t)n_1^H(s)\right\} p\left(t-iT\right)p\left(s-iT-\tau T\right) dt ds = 1 - \tau.
\end{align}

Thus, the covariance matrix of $\mathbf{N}_1$ in \eqref{eqU1receivedsampleX1+X2} is given by
\begin{equation}\label{eqRN=R}
\mathbf{R}_{\mathbf{N}_1} = \mathbb{E}\left\{\mathbf{N}_1\mathbf{N}_1^H\right\} = \mathbf{R}.
\end{equation}


Similarly, the received samples at User 2 in the broadcast phase can be written as
\begin{align}\label{eqY2broadcastmatrix}
\mathbf{Y}_{2} &= h_2\sqrt{P_1}\mathbf{RG}_1\mathbf{S_1} + h_2\sqrt{P_2}\mathbf{RG}_2\mathbf{S_2} + \mathbf{N}_2,
\end{align}
where the covariance matrix $\mathbf{R}_{\mathbf{N}_2} = \mathbb{E}\left\{\mathbf{N}_2\mathbf{N}_2^H\right\} = \mathbf{R}$.

\subsubsection{C-NOMA}
By setting $\tau = 0$, the C-ANOMA system becomes the C-NOMA system. For the C-NOMA systems, users do not use the oversampling technique. The $i$th sample at Users 1 and 2 in the broadcast phase will be
\begin{align}
y_{1}[i] &= h_1\sqrt{P_1}s_1[i] + h_1\sqrt{P_2}s_2[i] + n_1[i],\label{eqU1receivedsamplenoma}\\
y_{2}[i] &= h_2\sqrt{P_1}s_1[i] + h_2\sqrt{P_2}s_2[i] + n_2[i],\label{eqU2receivedsamplenoma}
\end{align}
where $n_j[i] = \int_{0}^{\infty}n_j(t)p(t-iT)dt$, $j = 1, 2$. Note that \eqref{eqU1receivedsamplenoma} and \eqref{eqU2receivedsamplenoma} can also be derived from \eqref{eqU1receivedsampleX1+X2} and \eqref{eqY2broadcastmatrix}, respectively, by letting $\tau = 0$.

\subsection{Relay Phase}
In the relay phase, User 2 receives another copy of the desired signal from User 1. The $i$th sample received at User 2 in the relay phase is given by
\begin{align}\label{eqy2relay}
y_{12}[i] = h_{12}\sqrt{P_r}s_2[i] + n_{12}[i],
\end{align}
where $P_r$ is the transmit power of User 1 and $n_{12}[i] = \int_{0}^{\infty}n_{12}(t)p(t-iT)dt$ is the additive noise. Note that the C-NOMA and C-ANOMA systems coincide in the relay phase. 

For ease of the following analysis, we rewrite the received samples from the relay link in \eqref{eqy2relay} into the matrix format, i.e.,
\begin{align}\label{eqY2relaymatrix}
\mathbf{Y}_{12} = h_{12}\sqrt{P_r}\mathbf{S}_2 + \mathbf{N}_{12},
\end{align}
where $\mathbf{Y}_{12} = [y_{12}[1],y_{12}[2],\cdots,y_{12}[N]]^T$, $\mathbf{N}_{12} = [n_{12}[1],n_{12}[2],\cdots,n_{12}[N]]^T$, and the covariance matrix $\mathbf{R}_{\mathbf{N}_{12}} = \mathbb{E}\left\{\mathbf{N}_{12}\mathbf{N}_{12}^H\right\} = \mathbf{I}_{N}$.

Combining all the received samples of User~2 in C-ANOMA systems, i.e., $\mathbf{Y}_{2}$ in \eqref{eqY2broadcastmatrix} and $\mathbf{Y}_{12}$ in \eqref{eqY2relaymatrix}, we have
\begin{align}\label{eqY2}
\mathbf{\tilde{Y}}_2 = \left[\begin{smallmatrix}
\mathbf{Y}_{2}\\ \mathbf{Y}_{12}
\end{smallmatrix}\right] = \underbrace{\left[\begin{smallmatrix}
h_2\sqrt{P_1}\mathbf{RG}_1\\ \mathbf{0}_{N}
\end{smallmatrix}\right]}_{\mathbf{W}_1} \mathbf{S}_1 + \underbrace{\left[\begin{smallmatrix}
h_2\sqrt{P_2}\mathbf{RG}_2\\h_{12}\sqrt{P_r}\mathbf{I}_{N}\end{smallmatrix}\right]}_{\mathbf{W}_2}\mathbf{S}_2 + \underbrace{\left[\begin{smallmatrix}
\mathbf{N}_2\\\mathbf{N}_{12}\end{smallmatrix}\right]}_{\mathbf{N}}.
\end{align}

Applying $\mathbb{E}\left\{\mathbf{N}_2\mathbf{N}_2^H\right\} = \mathbf{R}$ and $\mathbb{E}\left\{\mathbf{N}_{12}\mathbf{N}_{12}^H\right\} = \mathbf{I}_{N}$, the covariance matrix of the concatenated noise vector $\mathbf{N}$ is given by
\begin{align}
\mathbf{R}_{\mathbf{N}} = \mathbb{E}\left\{\mathbf{NN}^H\right\} = \left[\begin{smallmatrix}
\mathbb{E}\left\{\mathbf{N}_2\mathbf{N}_2^H\right\} &\mathbb{E}\left\{\mathbf{N}_2\mathbf{N}_{12}^H\right\}\\
\mathbb{E}\left\{\mathbf{N}_{12}\mathbf{N}_2^H\right\} &\mathbb{E}\left\{\mathbf{N}_{12}\mathbf{N}_{12}^H\right\}
\end{smallmatrix}\right] = \left[\begin{smallmatrix}
\mathbf{R} &\mathbf{0}\\
\mathbf{0} &\mathbf{I}
\end{smallmatrix}\right].
\end{align}

\section{Performance Analysis of C-ANOMA Systems}\label{secPerformanceAnalysis}
In this section, we analyze the individual throughput of users in the C-ANOMA and C-NOMA systems, including the strong and weak users. 

\subsection{Strong User}
\subsubsection{C-ANOMA}


In C-ANOMA systems, the block-wise SIC is adopted at User 1, i.e., it first decodes the block of symbols intended for User 2, subtracts it from the received signal, and then decodes the intended symbols. Note that BS transmits one block of symbols via two block times in the half-duplex mode~\cite{yue2018exploiting}. Besides, an extra $\tau$ time is utilized to create the sampling diversity in the symbol-asynchronous transmission. Hence, in the half-duplex C-ANOMA systems, a block of $N$ symbols are transmitted via $2N + \tau$ channel uses to Users~1 and 2. By considering \eqref{eqU1receivedsampleX1+X2} as a virtual multiple-input multiple-output (MIMO) system and treating the symbols for User 1 as noise, the throughput of User~1 to detect User~2's message is given by
\begin{align}\label{eqR2->1matrixform}
	R_{2\rightarrow 1}^{\mathrm{ANOMA}} &= \frac{1}{2N+\tau}\log\det\left[\mathbf{I}_{2N} + \left(\mathbf{R_{N_1}} + P_1|h_1|^2\mathbf{RG}_1\mathbf{G}_1^H\mathbf{R}^H\right)^{-1} P_2|h_1|^2\mathbf{RG}_2\mathbf{G}_2^H\mathbf{R}^H\right]\notag\\
	&\stackrel{(a)}{=} \frac{1}{2N+\tau}\log\det\left[\mathbf{I}_{2N} + \left(\mathbf{I}_{2N} + P_1|h_1|^2\mathbf{G}_1\mathbf{G}_1^H\mathbf{R}\right)^{-1} P_2|h_1|^2\mathbf{G}_2\mathbf{G}_2^H\mathbf{R}\right],
\end{align}
where $(a)$ is derived by applying $\mathbf{R_{N_1}} = \mathbf{R}$ and $\mathbf{R}^H = \mathbf{R}$.

Under the assumption of perfect SIC, by subtracting User 2's message from the superposed signal in \eqref{eqU1receivedsampleX1+X2}, the throughput of User 1 to detect its own message is calculated as
\begin{align}\label{eqR1matrixform}
R_1^{\mathrm{ANOMA}} &= \frac{1}{2N+\tau}\log\det\left(\mathbf{I}_{2N} + P_1|h_1|^2\mathbf{R}^{-1}_{\mathbf{N}_1}\mathbf{RG}_1\mathbf{G}_1^H\mathbf{R}^H\right)\notag\\
&= \frac{1}{2N+\tau}\log\det\left(\mathbf{I}_{2N} +  P_1|h_1|^2\mathbf{G}_1\mathbf{G}_1^H\mathbf{R}\right).
\end{align}

After matrix calculations, we can rewrite the throughput expressions at User 1 in \eqref{eqR2->1matrixform} and \eqref{eqR1matrixform} as functions of the receive signal-to-noise ratios (SNRs), i.e., $\mu_1$ and $\mu_2$, the normalized timing mismatch, $\tau$, and the frame length, $N$, i.e.,
\begin{align}
R_{2\rightarrow1}^{\mathrm{ANOMA}} &= \frac{1}{2N+\tau}\log \frac{\left(r_1^{N+1} - r_2^{N+1}\right) + \tau^2\left(r_1^N - r_2^N\right)}{r_1 - r_2} + \frac{N}{2N+\tau}\log\left(\frac{\mu_1\mu_2}{1 + \mu_1}\right),\label{eqR2->1analytical}\\
R_1^{\mathrm{ANOMA}} &= \frac{N}{2N+\tau}\log\left(1 + \mu_1\right),\label{eqR1analytical}
\end{align}
where 
\begin{align}
\mu_1 &= P_1|h_1|^2, \mu_2 = P_2|h_1|^2,Q=2\tau(1-\tau),\label{muexpression}\\
r_1 &=
\frac{\mu_1^{-1}\! +\! \mu_2^{-1}\! +\! \mu_1^{-1}\mu_2^{-1}\! +\! Q+ \sqrt{\left(\mu_1^{-1} + \mu_2^{-1} + \mu_1^{-1}\mu_2^{-1} + Q\right)^2\! -\! Q^2}}{2},\label{r1expression}\\
r_2 &=
\frac{\mu_1^{-1}\! +\! \mu_2^{-1}\! +\! \mu_1^{-1}\mu_2^{-1}\! +\! Q- \sqrt{\left(\mu_1^{-1} + \mu_2^{-1} + \mu_1^{-1}\mu_2^{-1} + Q\right)^2\! -\! Q^2}}{2}.\label{r2expression}
\end{align}
The detailed derivation of \eqref{eqR2->1analytical} and \eqref{eqR1analytical} is presented in Appendix~\ref{prooflemma(1+mu_1)N}.

\subsubsection{C-NOMA}
In conventional (synchronous) NOMA systems, with perfect SIC, the throughputs of User 1 are given by \cite{kim2015capacity,yue2018exploiting}
\begin{align}
R_{2\rightarrow1}^{\mathrm{NOMA}} &= \frac{1}{2}\log\left(1 + \frac{\mu_2}{1+\mu_1}\right),\label{eqR2->1NOMA}\\
R_1^{\mathrm{NOMA}} &= \frac{1}{2}\log(1 + \mu_1).\label{eqR1NOMA}
\end{align}

We note that by setting $\tau = 0$, we obtain $Q = 0$, $r_2 = 0$, and $r_1 = \mu_1^{-1} + \mu_2^{-1} + \mu_1^{-1}\mu_2^{-1}$. Thus, $R_{2\rightarrow1}^{\mathrm{ANOMA}}|_{\tau = 0} = R_{2\rightarrow1}^{\mathrm{NOMA}}$ and $R_{1}^{\mathrm{ANOMA}}|_{\tau = 0} = R_{1}^{\mathrm{NOMA}}$.

\subsubsection{Comparison between C-ANOMA and C-NOMA}
To study the throughput performance in the systems with a relatively large frame length, we consider the asymptotic case of $N\rightarrow \infty$. According to \eqref{eqR1analytical}, the throughput of User~1 to decode its own message if $N\rightarrow\infty$ is given by
\begin{align}
    R_{1,\mathrm{asymp}}^{\mathrm{ANOMA}} \stackrel{\triangle}{=} \lim_{N\rightarrow \infty} R_1^{\mathrm{ANOMA}} &= \frac{1}{2}\log\left(1 + \mu_1\right) = R_1^{\mathrm{NOMA}}.\label{eqR1asympANOMA}
\end{align}	

We note from \eqref{eqR1asympANOMA} that User~1 in C-ANOMA and C-NOMA systems can achieve the same throughput to detect its own message for a sufficiently large frame length. It is because with perfect SIC, the throughput of User 1 to detect its own message is not affected by the symbol asynchrony of the signal for User~2. Furthermore, we derive the following theorem to compare the throughputs of User~1 to detect User~2's message in the C-ANOMA and C-NOMA systems.

\begin{Theorem}\label{ThmR2->1ANOMA>R2->1NOMA}
	The throughputs of User 1 to detect User 2's message in the C-NOMA and C-ANOMA systems satisfy the following inequalities
	\begin{align}
	R_{2\rightarrow1}^{\mathrm{NOMA}} &\leq R_{2\rightarrow1,L}^{\mathrm{ANOMA}} \stackrel{\triangle}{=} \frac{1}{2}\log\left(1+\frac{\mu_2+\frac{1}{2}\mu_1\mu_2Q}{1+ \mu_1}\right)\notag\\
	&\leq R_{2\rightarrow1,\mathrm{asymp}}^{\mathrm{ANOMA}} \stackrel{\triangle}{=} \lim_{N\rightarrow \infty} R_{2\rightarrow1}^{\mathrm{ANOMA}} = \frac{1}{2}\log\left(\frac{\mu_1\mu_2 r_1}{1 + \mu_1}\right)\label{eqR2->1asympANOMA}\\
	&\leq R_{2\rightarrow1,U}^{\mathrm{ANOMA}} \stackrel{\triangle}{=} \frac{1}{2}\log\left(1+\frac{\mu_2+\mu_1\mu_2Q}{1+ \mu_1}\right),\notag
	\end{align}
	where $Q = 2\tau(1-\tau)$, all the equal signs are achieved if and only if $\tau = 0$.
\end{Theorem}
\begin{IEEEproof}
	See Appendix \ref{proofThmR2ANOMA>R2NOMA}.
\end{IEEEproof}

We note from Theorem \ref{ThmR2->1ANOMA>R2->1NOMA} that for a relatively large frame length, User 1 in C-ANOMA systems can achieve a higher throughput to decode User 2's message compared with that in C-NOMA systems. Besides, comparing the expressions for $R_{2\rightarrow1,L}^{\mathrm{ANOMA}}$ and $R_{2\rightarrow1}^{\mathrm{NOMA}}$, we find that the gain of C-ANOMA systems is related to the term $\mu_1\mu_2Q$ which increases as the channel qualities improve. 

In practice, the frame length, $N$, is determined by several factors, such as the channel coherence time, the modulation, the sampling rate, etc., which is beyond the scope of this paper. We assume that the frame length $N$ is a predetermined parameter in this paper. We will show in Section~\ref{sec:numericalResults} that the asymptotic throughput well approximates the actual one for not-so-large values of $N$, e.g, $N > 50$.

\subsection{Weak User}
In the half-duplex cooperative relaying scenario, the weak user, User 2, receives two blocks of symbols, one from BS with the superposed signal through the broadcast link and the other one from User 1 with only the intended signal through the relay link. 
\subsubsection{C-ANOMA} 
Treating \eqref{eqY2} as a virtual MIMO system and considering User 1's message as noise, the throughput of User 2 can be calculated as
\begin{align}\label{eqR2ANOMAmatrix}
R_2^{\mathrm{ANOMA}} = \frac{1}{2N+\tau}\log\det\left[\mathbf{I}_{3N} + \left(\mathbf{R_N} + \mathbf{W}_1\mathbf{W}_1^H\right)^{-1}\mathbf{W}_2\mathbf{W}_2^H\right].
\end{align}

The throughput of User 2 can be written as a function of the transmit powers, the channel gains, the normalized timing mismatch, and the frame length in the following theorem.
\begin{Theorem}\label{theoremR2ANOMA}
	In the half-duplex C-ANOMA systems, the throughput of User 2 is given by
	\begin{align}\label{eqR2ANOMA}
	R_2^{\mathrm{ANOMA}} &= \frac{1}{2N+\tau}\log \frac{\left(z_1^{N+1} - z_2^{N+1}\right) + \tau^2\left(z_1^N - z_2^N\right)}{z_1 - z_2} + \frac{N}{2N+\tau}\log\left(\frac{P_1P_2|h_{2}|^4}{1 + P_1|h_2|^2}\right),
	\end{align}
	where
	\begin{align}
	\nu_1 &= P_1|h_2|^2, \nu_2 = \frac{P_2|h_2|^2}{1+P_r|h_{12}|^2},Q=2\tau(1-\tau)\label{eqnuexpression}\\
	z_1 &=
	\frac{\nu_1^{-1}\! +\! \nu_2^{-1}\! +\! \nu_1^{-1}\nu_2^{-1}\! +\! Q+ \sqrt{\left[\nu_1^{-1} + \nu_2^{-1} + \nu_1^{-1}\nu_2^{-1} + Q\right]^2\! -\! Q^2}}{2},\label{eqz1expression}\\
	z_2 &=
	\frac{\nu_1^{-1}\! +\! \nu_2^{-1}\! +\! \nu_1^{-1}\nu_2^{-1}\! +\! Q- \sqrt{\left[\nu_1^{-1} + \nu_2^{-1} + \nu_1^{-1}\nu_2^{-1} + Q\right]^2\! -\! Q^2}}{2}.\label{eqz2expression}
	\end{align}
\end{Theorem}
\begin{IEEEproof}
	See Appendix \ref{prooftheoremR2ANOMA}.
\end{IEEEproof}

\subsubsection{C-NOMA}
In C-NOMA systems, User 2 adopts the maximal ratio combining (MRC) to combine the signals from the direct and relay links \cite{zhang2017full,yue2018exploiting}. Then, the throughput of User 2 is given by 
\begin{align}
R_2^{\mathrm{NOMA}} = \frac{1}{2}\log\left(1 + P_r|h_{12}|^2 + \frac{P_2|h_2|^2}{P_1|h_2|^2 + 1}\right).
\end{align}

Note that by setting $\tau = 0$, we have $Q = 0$, $z_2 = 0$, and $z_1 = \nu_1^{-1}\! +\! \nu_2^{-1}\! +\! \nu_1^{-1}\nu_2^{-1}$. Thus, the expression for the throughput of User~2 in C-ANOMA systems coincides with that in C-NOMA systems, i.e.,  $R_{2}^{\mathrm{ANOMA}}|_{\tau = 0} = R_{2}^{\mathrm{NOMA}}$.

\subsubsection{Comparion between C-ANOMA and C-NOMA}
We derive the following theorem which compares the throughputs of the C-ANOMA and C-NOMA systems for $N\rightarrow\infty$.
\begin{Theorem}\label{theoremR2ANOMA>R2NOMA}
In C-ANOMA systems, the throughput of User 2 for the asymptotic case of $N \rightarrow \infty$ is given by
\begin{align}\label{eqR2ANOMAasymp}
	R_{2,\mathrm{asymp}}^{\mathrm{ANOMA}} \stackrel{\triangle}{=}& \lim_{N\rightarrow \infty}\!R_{2}^{\mathrm{ANOMA}}\notag\\
	=&
    \frac{1}{2}\log\!\left[\frac{1 + P_r|h_{12}|^2}{2} + \frac{P_2|h_2|^2 + P_1P_2|h_2|^4Q}{2(1+ P_1|h_2|^2)}\right.\notag\\
	&+ \! \left. \!\frac{1}{2}\sqrt{\!\left(1 + P_r|h_{12}|^2 + \frac{ P_2|h_2|^2 \!+\! P_1P_2|h_2|^4Q }{1+ P_1|h_2|^2}\right)^2\!-\! \left(\frac{P_1P_2|h_2|^4Q}{1+ P_1|h_2|^2}\right)^2}\right],
\end{align}
where $Q = 2\tau(1-\tau)$.
	The throughputs of User 2 for the C-NOMA and C-ANOMA systems satisfy the following inequalities
	\begin{align}\label{eqR2ANOMA>R2NOMA}
	R_{2}^{\mathrm{NOMA}} &\leq R_{2,L}^{\mathrm{ANOMA}} \stackrel{\triangle}{=} \frac{1}{2}\log\left(1+P_r|h_{12}|^2+\frac{P_2|h_2|^2+\frac{1}{2}P_1P_2|h_2|^4Q}{1+ P_1|h_2|^2}\right)\notag\\
	&\leq R_{2,\mathrm{asymp}}^{\mathrm{ANOMA}}\notag\\
	&\leq R_{2,U}^{\mathrm{ANOMA}} \stackrel{\triangle}{=} \frac{1}{2}\log\left(1+P_r|h_{12}|^2+\frac{P_2|h_2|^2+P_1P_2|h_2|^4Q}{1+ P_1|h_2|^2}\right),
	\end{align}
	where the equal signs are achieved if and only if $\tau = 0$.
\end{Theorem}
\begin{IEEEproof}
See Appendix \ref{proofTheoremR2ANOMA>R2NOMA}.
\end{IEEEproof}

We note from \eqref{eqR2ANOMA>R2NOMA} that the gain of C-ANOMA over C-NOMA depends on the term $P_1P_2|h_2|^4Q$, thus, a better direct channel between User~2 and BS results in a greater performance improvement of C-ANOMA systems compared with C-NOMA systems. Moreover,
according to \eqref{eqR1asympANOMA}, Theorems \ref{ThmR2->1ANOMA>R2->1NOMA} and \ref{theoremR2ANOMA>R2NOMA}, it is shown that for $N\rightarrow \infty$, the throughputs of both users to detect the weak user's message in the C-ANOMA systems are larger than those in the C-NOMA systems while the throughput of the strong user to detect its own message is identical for the C-ANOMA and C-NOMA systems. In Section \ref{sec:numericalResults}, we show by numerical results that the C-ANOMA systems outperform the C-NOMA systems in terms of the throughputs to decode the weak user's message with a relatively small value of $N$, e.g., $N > 20$.

\section{C-ANOMA System Design}\label{sectionSysdesign}
In this section, we study the optimal design of the C-ANOMA systems, including the optimal timing mismatch and the power control strategy.
\subsection{Optimal Timing mismatch}
We first investigate the optimal normalized timing mismatch, $\tau^*$. Although the optimal normalized timing mismatch to maximize $R_{2\rightarrow 1}^{\mathrm{ANOMA}}$ and $R_2^{\mathrm{ANOMA}}$ is analytically intractable for a general finite frame length $N$, we can numerically obtain $\tau^*$ for a given finite $N$ by simply searching in the range of $0 \le \tau < 1$ as done in Section \ref{sec:numericalResults}. To derive the optimal $\tau$ for a large $N$, we study the asymptotic case of $N\rightarrow\infty$. According to \eqref{eqR1asympANOMA}, the throughput of User 1 to detect its own message is independent of $\tau$. According to \eqref{eqR2->1asympANOMA} and \eqref{eqR2ANOMAasymp}, it is easy to show that $R_{2\rightarrow1,\mathrm{asymp}}^{\mathrm{ANOMA}}$ and $R_{2,\mathrm{asymp}}^{\mathrm{ANOMA}}$ are increasing functions of $Q$ which is given by $2\tau(1-\tau)$. Thus, maximizing $R_{2\rightarrow1,\mathrm{asymp}}^{\mathrm{ANOMA}}$ and $R_{2,\mathrm{asymp}}^{\mathrm{ANOMA}}$ is equivalent to maximizing the term $\tau(1-\tau)$. Therefore, the optimal $\tau$ to maximize the throughputs of both users to detect User 2's message converges to 0.5, i.e.,
\begin{align}\label{eqtau*}
    \tau^* \stackrel{\triangle}{=} \mathop{\arg\max}_\tau R_{2\rightarrow1,\mathrm{asymp}}^{\mathrm{ANOMA}} = \mathop{\arg\max}_\tau R_{2,\mathrm{asymp}}^{\mathrm{ANOMA}} = 0.5. 
\end{align}

\subsection{Power Minimization}
In this paper, we consider the delay-tolerant mode where the BS and the relay user can dynamically adjust their transmit powers according to the channel states in order to avoid outage and satisfy the minimum rate requirements~\cite{yue2018exploiting}. Our objective is to minimize the weighted sum transmit power of BS and the relay user under the minimum rate (i.e., QoS) requirements and the individual power constraints. Then, the power minimization problem can be formulated as
\vspace{-8mm}
\begin{mini!}|s|[2]                   
	{_{P_1, P_2, P_r}}                               
	{\omega_s (P_1 + P_2) + \omega_r P_r, \label{optimal:eq1}}   
	{\label{eq:optimal}}             
	{}                                
	\addConstraint{R_{2\rightarrow 1}^{\mathrm{ANOMA}}}{\ge R_2^*,
		R_{1}^{\mathrm{ANOMA}}\ge R_1^*, R_{2}^{\mathrm{ANOMA}}\ge R_2^*\label{optimal:eqSatisfyTargetRateOriginal}}    
	\addConstraint{P_1 + P_2}{< P_{s,\max}, P_r< P_{r,\max},}  
\end{mini!}
\noindent where $\omega_s$ and $\omega_r$ are the non-negative weights for the transmit powers of BS and User 1, respectively, such that $\omega_s + \omega_r = 1$. $P_{s,\max}$ and $P_{r,\max}$ stand for the maximum available powers of BS and User 1, respectively. $R_1^*$ and $R_2^*$ are the target rates of Users 1 and 2's messages. Note that the choice of $\omega_s$ and $\omega_r$ provides a trade-off between the power consumption of BS and that of the relay user. For instance, if one wants to further restrict the power consumption of the relay user due to its limited battery capacity, $\omega_r$ should be chosen greater than $\omega_s$. 

The exact expressions of $R_{2\rightarrow 1}^{\mathrm{ANOMA}}$ and $R_{2}^{\mathrm{ANOMA}}$ in \eqref{eqR2->1analytical} and \eqref{eqR2ANOMA} make the optimization problem \eqref{eq:optimal} analytically intractable. To simplify the optimization problem, we replace $R_{2\rightarrow 1}^{\mathrm{ANOMA}}$ and $R_{2}^{\mathrm{ANOMA}}$ in \eqref{optimal:eqSatisfyTargetRateOriginal}
with their \emph{asymptotic lower bounds}, which can provide a suboptimal solution for the original optimization problem \eqref{eq:optimal}, i.e.,
\begin{mini!}|s|[2]                   
	{_{P_1, P_2, P_r}}                               
	{\omega_s (P_1 + P_2) + \omega_r P_r, \label{suboptimal:eq1}}   
	{\label{eq:suboptimal}}             
	{}                                
	\addConstraint{R_{2\rightarrow 1,L}^{\mathrm{ANOMA}}}{\ge R_2^*\label{suboptimal:eqR2->1>R2*}}    
	\addConstraint{R_{1}^{\mathrm{ANOMA}}}{\ge R_1^*\label{suboptimal:eqR1>R1*}}  
	\addConstraint{R_{2,L}^{\mathrm{ANOMA}}}{\ge R_2^*\label{suboptimal:eqR2>R2*}}
	\addConstraint{P_1 + P_2}{< P_{s,\max}, P_r < P_{r,\max}.\label{suboptimal:eqRangesAlphaPsPr}}
\end{mini!} 
\indent For sufficiently large values of $N$, Eqs. \eqref{suboptimal:eqR2->1>R2*} and \eqref{suboptimal:eqR2>R2*} are stronger constraints for $R_{2\rightarrow 1}^{\mathrm{ANOMA}}$ and $R_{2}^{\mathrm{ANOMA}}$ compared with those in \eqref{optimal:eqSatisfyTargetRateOriginal}, which means that the solution of \eqref{eq:optimal} can do at least as good as that of \eqref{eq:suboptimal}. In what follows, we explain that \eqref{eq:suboptimal} can also provide a suboptimal solution of \eqref{eq:optimal} for a finite $N$. By definition, as $N$ increases, the exact throughputs can be arbitrarily close to the asymptotic ones. We assume that $R_{2\rightarrow 1}^{\mathrm{ANOMA}} \ge R_{2\rightarrow 1,L}^{\mathrm{ANOMA}}$ for any $N \ge N_1$ and $R_{2}^{\mathrm{ANOMA}} \ge R_{2,L}^{\mathrm{ANOMA}}$ for any $N \ge N_2$. By choosing a proper $N^*$, for example, $N^* = \max\{N_1, N_2\}$, we can ensure that $R_{2\rightarrow 1}^{\mathrm{ANOMA}} \ge R_{2\rightarrow 1,L}^{\mathrm{ANOMA}}$ and $R_{2}^{\mathrm{ANOMA}} \ge R_{2,L}^{\mathrm{ANOMA}}$ for the given $N^*$. We will show that $N^*$ can be a reasonable value (e.g., $N^* = 100$) in the numerical results section. In practice, the actual frame length is usually greater than 100. For example, in global system for mobile communications (GSM), there are approximately 156 symbols in a normal burst (a physical channel carrying information on traffic and control channels) \cite{gibson2012mobile}. As a result, the optimization problem \eqref{eq:suboptimal} can provide a suboptimal solution for the problem \eqref{eq:optimal} with the frame length used in practical communication systems.

By simplifying \eqref{suboptimal:eqR2->1>R2*}, \eqref{suboptimal:eqR1>R1*}, and \eqref{suboptimal:eqR2>R2*}, we obtain
\begin{align}
P_2 &\ge \frac{\gamma_2}{|h_1|^2}\frac{1+P_1|h_1|^2}{1 + \frac{1}{2}QP_1|h_1|^2},\label{eqP2conditionconvert}\\
P_1 &\ge \frac{\gamma_1 + \epsilon}{|h_1|^2},\label{eqP1conditionconvert}\\
P_r &\ge \frac{\gamma_2}{|h_{12}|^2} - \frac{P_2|h_2|^2}{|h_{12}|^2}\frac{1+\frac{1}{2}QP_1|h_2|^2}{1+P_1|h_2|^2}.\label{eqPrconditionconvert}
\end{align}
where $\gamma_i = 2^{2R_i^*} - 1$, $i = 1, 2$, is the target signal-to-interference-plus-noise ratio (SINR) to detect User~$i$'s message, $\epsilon = 2^{2R_1^*}(2^{\frac{\tau}{N}R_1^*} - 1)$, and $Q = 2\tau(1-\tau)$. The value of $\epsilon$ can be made arbitrary small with increasing $N$. For a sufficiently large $N$, i.e., $N > N^*$, we have $\epsilon < \epsilon^* \stackrel{\triangle}{=} 2^{2R_1^*}(2^{\frac{\tau}{N^*}R_1^*} - 1)$, hence, we can substitute  \eqref{eqP1conditionconvert} with a stronger constraint, i.e.,
\begin{align}
P_1 &\ge \frac{\gamma_1 + \epsilon^*}{|h_1|^2}.\label{eqP1conditionconvert2}
\end{align}

Then, by replacing the constraints with \eqref{eqP2conditionconvert}, \eqref{eqPrconditionconvert}, and \eqref{eqP1conditionconvert2}, the optimization problem \eqref{eq:suboptimal} becomes
\vspace{-5mm}
\begin{mini!}|s|[2]                   
	{_{P_1, P_2, P_r}}                               
	{\omega_s (P_1 + P_2) + \omega_r P_r, \label{suboptimal2:eq1}}   
	{\label{eq:suboptimal2}}             
	{}                                
	\addConstraint{\frac{\gamma_1+\epsilon^*}{|h_1|^2}}{\le P_1 \le P_{s,\max}\label{suboptimal2:eqP1constraint}}    
	\addConstraint{ \frac{\gamma_2}{|h_1|^2}\frac{1+P_1|h_1|^2}{1 + \frac{1}{2}QP_1|h_1|^2} \le P_2 \le P_{s,\max} - P_1\label{suboptimal2:eqP2constraint}}  
	\addConstraint{\zeta_r}{\stackrel{\triangle}{=} \max\left\{0, \frac{\gamma_2}{|h_{12}|^2} - \frac{P_2|h_2|^2}{|h_{12}|^2}\frac{1+\frac{1}{2}QP_1|h_2|^2}{1+P_1|h_2|^2}\right\} \le P_r \le P_{r,\max}.\label{suboptimal2:eqPrconstraint}}
\end{mini!}
\indent Note that \eqref{suboptimal2:eqPrconstraint} indicates that the feasible domain of $P_r$ depends on $P_1$ and $P_2$ while the constraints of $P_1$ and $P_2$ in \eqref{suboptimal2:eqP1constraint} and \eqref{suboptimal2:eqP2constraint} do not rely on $P_r$. For any given $P_1$ and $P_2$, the weighted sum power is minimized when $P_r$ is equal to the least possible value, i.e., $P_r = \zeta_r$. Besides, we note that increasing $P_1$ improves $R_1^{\mathrm{ANOMA}}$ while worsens $R_{2, L}^{\mathrm{ANOMA}}$ and $R_{2\rightarrow 1, L}^{\mathrm{ANOMA}}$ due to the increased interference from User~1's message. Then, the powers $P_2$ and $P_r$ have to increase to counteract the interference of User 1's message. As a result, $P_1$ should also be chosen as the least possible value within the feasible domain \eqref{suboptimal2:eqP1constraint} to minimize the weighted sum power, i.e., $P_1 = \frac{\gamma_1+\epsilon^*}{|h_1|^2}$. By substituting the values of $P_r$ and $P_1$, the optimization problem \eqref{eq:suboptimal2} becomes
\vspace{-5mm}
\begin{mini!}|s|[2]                   
	{_{P_2}}                               
	{\omega_s P_2 + \omega_r \max\left\{0, \frac{\gamma_2}{|h_{12}|^2} - \frac{P_2|h_2|^2}{|h_{12}|^2}\frac{|h_1|^2+\frac{1}{2}Q(\gamma_1+\epsilon^*)|h_2|^2}{|h_1|^2+(\gamma_1+\epsilon^*)|h_2|^2}\right\}, \label{suboptimal3:eq1}}   
	{\label{eq:suboptimal3}}             
	{}                                
	\addConstraint{\zeta_2}{\le P_2 \le P_{s,\max} - \frac{\gamma_1+\epsilon^*}{|h_1|^2},\label{suboptimal3:constraint}}    
\end{mini!}
where
\begin{align}\label{eqzeta2exp}
\zeta_2 = \max\left\{\frac{\gamma_2}{|h_1|^2}\frac{1+\gamma_1+\epsilon^*}{1 + \frac{1}{2}Q(\gamma_1+\epsilon^*)}, \left(\frac{\gamma_2}{|h_{12}|^2} - P_{r,\max}\right)\frac{|h_{12}|^2}{|h_2|^2}\frac{|h_1|^2+(\gamma_1+\epsilon^*)|h_2|^2}{|h_1|^2+\frac{1}{2}Q(\gamma_1+\epsilon^*)|h_2|^2}\right\},
\end{align}
and the rightmost term in \eqref{eqzeta2exp} is derived by setting $\zeta_r < P_{r,\max}$.

We note from \eqref{suboptimal2:eqP1constraint} and \eqref{suboptimal3:constraint} that if $\frac{\gamma_1+\epsilon^*}{|h_1|^2} > P_{s,\max}$ or $\zeta_2 > P_{s,\max} - \frac{\gamma_1+\epsilon^*}{|h_1|^2}$, there is no valid solution for the power minimization problem, i.e., QoS cannot be satisfied with the limited transmit powers of BS and the relay user. The following analysis is under the assumption that there are valid solutions for the power minimization problem. 

By setting $\frac{\gamma_2}{|h_{12}|^2} - \frac{P_2|h_2|^2}{|h_{12}|^2}\frac{|h_1|^2+\frac{1}{2}Q(\gamma_1+\epsilon^*)|h_2|^2}{|h_1|^2+(\gamma_1+\epsilon^*)|h_2|^2} = 0$, we obtain
\begin{align}\label{eqzeta2*}
P_2 = \zeta_2^* \stackrel{\triangle}{=} \frac{\gamma_2}{|h_2|^2}\frac{|h_1|^2+(\gamma_1+\epsilon^*)|h_2|^2}{|h_1|^2+\frac{1}{2}Q(\gamma_1+\epsilon^*)|h_2|^2}.
\end{align}

\begin{figure}[!ht b]
	\centering
	\includegraphics[width=5in]{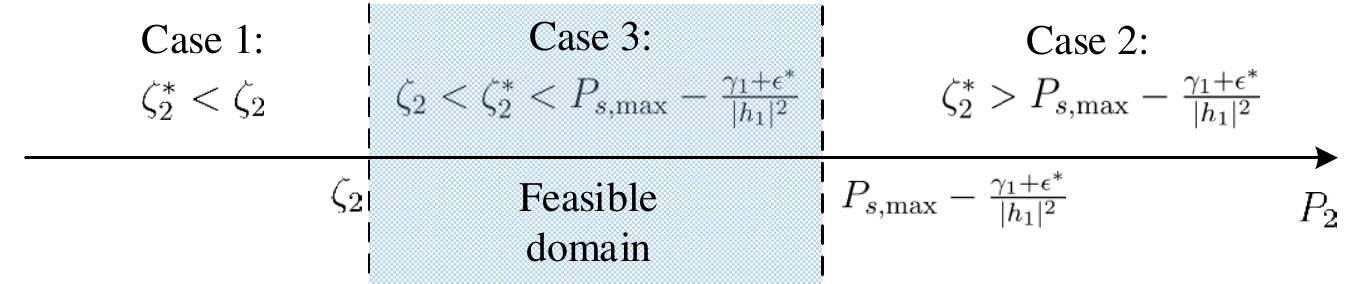}
	\caption{Illustration of the relationship among $\zeta_2^*$, $\zeta_2$, and $P_{s,\max} - \frac{\gamma_1+\epsilon^*}{|h_1|^2}$.}
	\label{fig_illustrate_3_cases}
\end{figure}

Then, the objective function \eqref{suboptimal3:eq1} becomes $\omega_s P_2 + \omega_r \left(\frac{\gamma_2}{|h_{12}|^2} - \frac{P_2|h_2|^2}{|h_{12}|^2}\frac{|h_1|^2+\frac{1}{2}Q(\gamma_1+\epsilon^*)|h_2|^2}{|h_1|^2+(\gamma_1+\epsilon^*)|h_2|^2}\right)$ if $P_2 < \zeta_2^*$ and $\omega_sP_2$ otherwise. As shown in Fig. \ref{fig_illustrate_3_cases}, we separate the following analysis into three cases according to the relationship among $\zeta_2^*$, $\zeta_2$, and $P_{s,\max}-\frac{\gamma_1+\epsilon^*}{|h_1|^2}$.

\subsubsection{Case 1}
If $\zeta_2^* < \zeta_2$, the optimization problem \eqref{eq:suboptimal3} becomes
\begin{align}\label{eq:suboptimal3case1}
\min_{P_2} P_2,\ \ \mathrm{s.t.}\ \zeta_2 \le P_2 \le P_{s,\max} - \frac{\gamma_1+\epsilon^*}{|h_1|^2},
\end{align}
\indent In this case, it is easy to obtain that the optimal transmit powers are $P_1^* = \frac{\gamma_1+\epsilon^*}{|h_1|^2}$, $P_2^* = \zeta_2$, and $P_r^* = 0$. 
Intuitively, this case indicates that the channel between BS and User 2 is strong enough such that no relay transmission is needed to satisfy the QoS at User 2.

\subsubsection{Case 2}
If $\zeta_2^* > P_{s,\max} - \frac{\gamma_1+\epsilon^*}{|h_1|^2}$, the optimization problem \eqref{eq:suboptimal3} becomes
\begin{mini!}|s|[2]                   
	{_{P_2}}                               
	{\omega_s P_2 + \omega_r \left(\frac{\gamma_2}{|h_{12}|^2} - \frac{P_2|h_2|^2}{|h_{12}|^2}\frac{|h_1|^2+\frac{1}{2}Q(\gamma_1+\epsilon^*)|h_2|^2}{|h_1|^2+(\gamma_1+\epsilon^*)|h_2|^2}\right), \label{suboptimal3case2:eq1}}   
	{\label{eq:suboptimal3case2}}             
	{}                                
	\addConstraint{\zeta_2}{\le P_2 \le P_{s,\max} - \frac{\gamma_1+\epsilon^*}{|h_1|^2}.}    
\end{mini!}
\indent By omitting the constant terms, the objective function \eqref{suboptimal3case2:eq1} becomes 
\begin{align}
\min_{P_2}\ \left(\omega_s - \omega_r \frac{|h_2|^2}{|h_{12}|^2}\frac{|h_1|^2+\frac{1}{2}Q(\gamma_1+\epsilon^*)|h_2|^2}{|h_1|^2+(\gamma_1+\epsilon^*)|h_2|^2}\right)P_2.
\end{align}

Note that the solution of \eqref{eq:suboptimal3case2} depends on the values of $\omega_s$ and $\omega_r$. The solutions can be given as follows: If $\omega_s = \omega_r \frac{|h_2|^2}{|h_{12}|^2}\frac{|h_1|^2+\frac{1}{2}Q(\gamma_1+\epsilon^*)|h_2|^2}{|h_1|^2+(\gamma_1+\epsilon^*)|h_2|^2}$, the optimal transmit powers are
\begin{align}
P_1^* &= \frac{\gamma_1+\epsilon^*}{|h_1|^2}, P_2^* \in \left[\zeta_2, P_{s,\max} - \frac{\gamma_1+\epsilon^*}{|h_1|^2}\right], P_r^* = \frac{\gamma_2}{|h_{12}|^2} - \frac{P_2^*|h_2|^2}{|h_{12}|^2}\frac{|h_1|^2+\frac{1}{2}Q(\gamma_1+\epsilon^*)|h_2|^2}{|h_1|^2+(\gamma_1+\epsilon^*)|h_2|^2},\label{eqPrIntuitiveExplain}
\end{align}
where $P_2^*$ can be any value in the given range. We provide the intuitive explanation for the solution as follows: One can observe from \eqref{eqPrIntuitiveExplain} that $P_r^*$ decreases with $P_2^*$. Under certain conditions on $\omega_r$ and $\omega_s$, i.e., $\omega_s = \omega_r \frac{|h_2|^2}{|h_{12}|^2}\frac{|h_1|^2+\frac{1}{2}Q(\gamma_1+\epsilon^*)|h_2|^2}{|h_1|^2+(\gamma_1+\epsilon^*)|h_2|^2}$, the weighted sum power will be constant as the decrease in $P_r^*$ is equal to the increase in $P_2^*$.

If $\omega_s > \omega_r \frac{|h_2|^2}{|h_{12}|^2}\frac{|h_1|^2+\frac{1}{2}Q(\gamma_1+\epsilon^*)|h_2|^2}{|h_1|^2+(\gamma_1+\epsilon^*)|h_2|^2}$, $P_2^*$ should be chosen the least possible value. The optimal transmit powers are given by
\begin{align}
P_1^* &= \frac{\gamma_1+\epsilon^*}{|h_1|^2},
P_2^* = \zeta_2,
P_r^* = \frac{\gamma_2}{|h_{12}|^2} - \frac{P_2^*|h_2|^2}{|h_{12}|^2}\frac{|h_1|^2+\frac{1}{2}Q(\gamma_1+\epsilon^*)|h_2|^2}{|h_1|^2+(\gamma_1+\epsilon^*)|h_2|^2}.
\end{align}

If $\omega_s < \omega_r \frac{|h_2|^2}{|h_{12}|^2}\frac{|h_1|^2+\frac{1}{2}Q(\gamma_1+\epsilon^*)|h_2|^2}{|h_1|^2+(\gamma_1+\epsilon^*)|h_2|^2}$, $P_2^*$ should choose the largest possible value. The optimal transmit powers are given by
\begin{align}\label{eqOptimalPsForANOMANOMAcomparison}
P_1^* &= \frac{\gamma_1+\epsilon^*}{|h_1|^2},
P_2^* = P_{s,\max} - \frac{\gamma_1+\epsilon^*}{|h_1|^2},
P_r^* = \frac{\gamma_2}{|h_{12}|^2} - \frac{P_2^*|h_2|^2}{|h_{12}|^2}\frac{|h_1|^2+\frac{1}{2}Q(\gamma_1+\epsilon^*)|h_2|^2}{|h_1|^2+(\gamma_1+\epsilon^*)|h_2|^2}.
\end{align}

\subsubsection{Case 3}
If $\zeta_2 < \zeta_2^* < P_{s,\max} - \frac{\gamma_1+\epsilon^*}{|h_1|^2}$, the optimization problem \eqref{eq:suboptimal3} becomes two sub-problems, i.e.,
\begin{align}\label{eq:suboptimal3case3_1}
\min_{P_2} P_2,\ \ \mathrm{s. t.}\ \zeta_2^*\le P_2 \le P_{s,\max} - \frac{\gamma_1+\epsilon^*}{|h_1|^2},
\end{align}
and
\begin{align}\label{eq:suboptimal3case3_2}
\min_{P_2} P_2\left(\omega_s - \omega_r \frac{P_2|h_2|^2}{|h_{12}|^2}\frac{|h_1|^2+\frac{1}{2}Q(\gamma_1+\epsilon^*)|h_2|^2}{|h_1|^2+(\gamma_1+\epsilon^*)|h_2|^2}\right),\ \ \mathrm{s.t.}\ \zeta_2\le P_2 \le \zeta_2^*.
\end{align}
\indent By following the derivation of Case 1, one can solve the problem \eqref{eq:suboptimal3case3_1}. Similarly, by following the steps of Case 2, one can solve the problem \eqref{eq:suboptimal3case3_2}. We assume that the optimal transmit powers for \eqref{eq:suboptimal3case3_1} and \eqref{eq:suboptimal3case3_2} are [$\tilde{P_1}$, $\tilde{P_2}$, $\tilde{P_r}$] and [$\bar{P_1}$, $\bar{P_2}$, $\bar{P_r}$], respectively. Then, the solution for Case 3 is given by
\begin{align}
[P_1^*, P_2^*, P_r^*] = \mathop{\arg\min}_{[P_1,P_2,P_r]\in\left\{[\tilde{P_1}, \tilde{P_2}, \tilde{P_r}], [\bar{P_1}, \bar{P_2}, \bar{P_r}]\right\}} \omega_s(P_1 + P_2) + \omega_r P_r.
\end{align}

To summarize the solutions, we provide Algorithm 1 to solve the problem \eqref{eq:suboptimal2}.
\begin{algorithm}[!h]
 \caption{Algorithm to find the optimal powers under QoS constraints}
 \begin{algorithmic}[1]
 \Function{Solve\_case\_1}{$L$, $U$}
 \State \Return $P_2^* = L$, $P_r^* = 0$.
 \EndFunction
 \Function{Solve\_case\_2}{$L$, $U$}
 \State \textbf{if} $\omega_s = \omega_r \frac{|h_2|^2}{|h_{12}|^2}\frac{|h_1|^2+\frac{1}{2}Q(\gamma_1+\epsilon^*)|h_2|^2}{|h_1|^2+(\gamma_1+\epsilon^*)|h_2|^2}$, \textbf{then} $P_2^* = \mathrm{random}\left( \left[\zeta_2, P_{s,\max} - \frac{\gamma_1+\epsilon^*}{|h_1|^2}\right]\right)$.
 \State \textbf{else if} $\omega_s > \omega_r \frac{|h_2|^2}{|h_{12}|^2}\frac{|h_1|^2+\frac{1}{2}Q(\gamma_1+\epsilon^*)|h_2|^2}{|h_1|^2+(\gamma_1+\epsilon^*)|h_2|^2}$, \textbf{then} $P_2^* = L$.
 \State \textbf{else} $P_2^* = U$.
 \State \Return $P_2^*$, $P_r^* = \frac{\gamma_2}{|h_{12}|^2} - \frac{P_2^*|h_2|^2}{|h_{12}|^2}\frac{|h_1|^2+\frac{1}{2}Q(\gamma_1+\epsilon^*)|h_2|^2}{|h_1|^2+(\gamma_1+\epsilon^*)|h_2|^2}$.
 \EndFunction
  \State \textbf{if} {$\frac{\gamma_1+\epsilon^*}{|h_1|^2} > P_{s,\max}$ or $\zeta_2 > P_{s,\max} - \frac{\gamma_1+\epsilon^*}{|h_1|^2}$,} \textbf{then} there is no solution, break.
  \State $P_1^* = \frac{\gamma_1 + \epsilon^*}{|h_1|^2}$
  \State \textbf{if} {$\zeta_2^* < \zeta_2$,} \textbf{then} $P_2^*$, $P_r^*$ = \textproc{Solve\_case\_1}($\zeta_2$, $P_{s,\max} - \frac{\gamma_1 + \epsilon^*}{|h_1|^2}$).
  \State \textbf{else if} {$\zeta_2^* > P_{s,\max} - \frac{\gamma_1+\epsilon^*}{|h_1|^2}$,} \textbf{then} $P_2^*$, $P_r^*$ = \textproc{Solve\_case\_2}($\zeta_2$, $P_{s,\max} - \frac{\gamma_1 + \epsilon^*}{|h_1|^2}$).
  \State \textbf{else}
  \State \indent $\tilde{P_2^*}, \tilde{P_r^*} = \textproc{Solve\_case\_1}(\zeta_2^*, P_{s,\max} - \frac{\gamma_1 + \epsilon^*}{|h_1|^2})$.
  \State \indent $\hat{P_2^*}, \hat{P_r^*} = \textproc{Solve\_case\_2}(\zeta_2, \zeta_2^*)$.
  \State \indent $P_2^*, P_r^* = \mathop{\arg\min}_{[P_2, P_r] \in \{[\tilde{P_2^*}, \tilde{P_r^*}], [\hat{P_2^*}, \hat{P_r^*}]\}} \omega_s P_2 + \omega_r P_r$.
 \State \Return $P_1^*$, $P_2^*$, $P_r^*$.
 \end{algorithmic}
 \end{algorithm}

\subsection{Comparison with C-NOMA}
According to \eqref{eqR1asympANOMA}, \eqref{eqR2->1asympANOMA}, Theorems \ref{ThmR2->1ANOMA>R2->1NOMA} and \ref{theoremR2ANOMA>R2NOMA}, the expressions for the throughputs in C-ANOMA systems, $R_{2\rightarrow1,L}^{\mathrm{ANOMA}}$, $R_{1}^{\mathrm{ANOMA}}$, and $R_{2,L}^{\mathrm{ANOMA}}$, become those in C-NOMA systems, $R_{2\rightarrow1}^{\mathrm{NOMA}}$, $R_1^{\mathrm{NOMA}}$, and $R_{2}^{\mathrm{NOMA}}$, by setting $\tau = 0$. Therefore, the solutions derived in the previous subsection can be applied to the C-NOMA systems simply by setting $\tau = 0$ which then results in $\epsilon^* = 0$ and $Q = 0$. For the C-NOMA systems, the power minimization problem \eqref{eq:suboptimal2} becomes
\begin{mini!}|s|[2]                   
	{_{P_1, P_2, P_r}}                               
	{\omega_s (P_1 + P_2) + \omega_r P_r, \label{suboptimal2NOMA:eq1}}   
	{\label{eq:suboptimal2NOMA}}             
	{}                                
	\addConstraint{\frac{\gamma_1}{|h_1|^2}}{\le P_1 \le P_{s,\max}\label{suboptimal2NOMA:eqP1constraint}}    
\addConstraint{\frac{\gamma_2\left(1+P_1|h_1|^2\right)}{|h_1|^2} \le P_2 \le P_{s,\max} - P_1\label{suboptimal2NOMA:eqP2constraint}}  
	\addConstraint{\max\left\{0, \frac{\gamma_2}{|h_{12}|^2} - \frac{P_2|h_2|^2}{|h_{12}|^2\left(1+P_1|h_2|^2\right)}\right\} \le P_r \le P_{r,\max}.\label{suboptimal2NOMA:eqPrconstraint}}
\end{mini!}
\indent Note that the feasible domains of $P_2$ and $P_r$ in \eqref{suboptimal2NOMA:eqP2constraint} and \eqref{suboptimal2NOMA:eqPrconstraint} are the subsets of those in \eqref{suboptimal2:eqP2constraint} and \eqref{suboptimal2:eqPrconstraint}, respectively. As a result, for a sufficiently large $N$, the minimization problem \eqref{eq:suboptimal2} for the C-ANOMA systems is a relaxation of the minimization problem \eqref{eq:suboptimal2NOMA} for the C-NOMA systems~\cite{geoffrion1971duality}. That is, the problem \eqref{eq:suboptimal2} provides a solution to minimize the weighted sum power within a wider feasible domain compared with \eqref{eq:suboptimal2NOMA}. In other words, if [$P_{1,\mathrm{ANOMA}}^*$, $P_{2,\mathrm{ANOMA}}^*$, $P_{r,\mathrm{ANOMA}}^*$] and [$P_{1,\mathrm{NOMA}}^*$, $P_{2,\mathrm{NOMA}}^*$, $P_{r,\mathrm{NOMA}}^*$] are the optimal solutions for \eqref{eq:suboptimal2} and \eqref{eq:suboptimal2NOMA}, respectively, we have
\begin{align}
\omega_s(P_{1,\mathrm{ANOMA}}^* + P_{2,\mathrm{ANOMA}}^*) + \omega_r P_{r,\mathrm{ANOMA}}^* \le \omega_s(P_{1,\mathrm{NOMA}}^* + P_{2,\mathrm{NOMA}}^*) + \omega_r P_{r,\mathrm{NOMA}}^*.\label{eqPsumANOMA<PsumNOMA}
\end{align}

We note from \eqref{eqPsumANOMA<PsumNOMA} that for a sufficiently large frame length, the C-ANOMA systems can consume less power compared with the C-NOMA systems in order to guarantee the same QoS. We will illustrate this phenomenon with numerical results in Section \ref{sec:numericalResults}.

\begin{figure}[!ht b]
	\centering
	\includegraphics[width=6.5in]{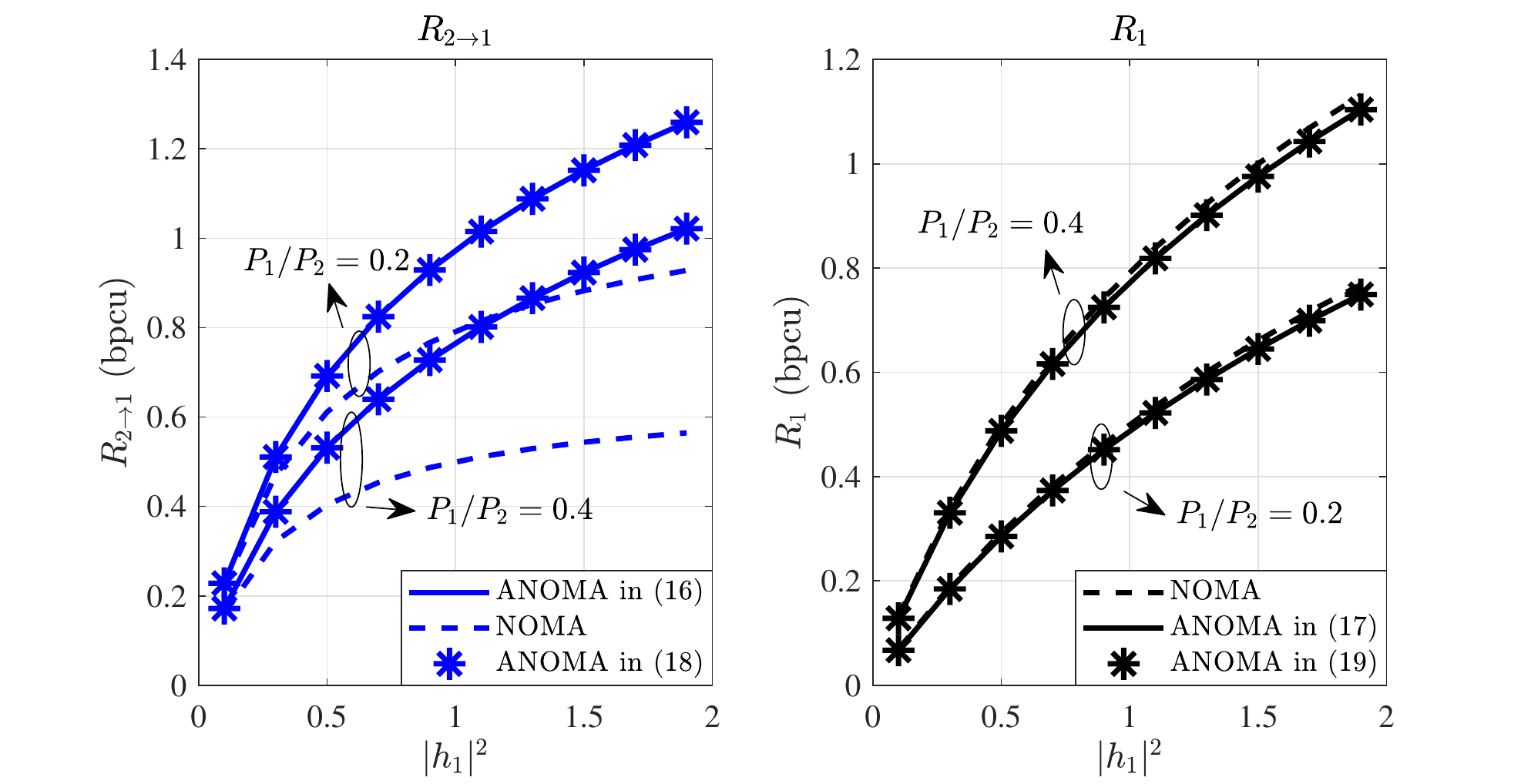}
	\caption{The throughputs $R_{2\rightarrow 1}$ and $R_1$ as functions of the channel gain $|h_1|^2$ for C-ANOMA and C-NOMA systems when $N = 10$, $\tau = 0.5$, $P_1 + P_2 = 5$.}
	\label{fig_rate_vs_h_R2-1}
\end{figure}
\begin{figure}[!ht b]
	\centering
	\includegraphics[width=4.5in]{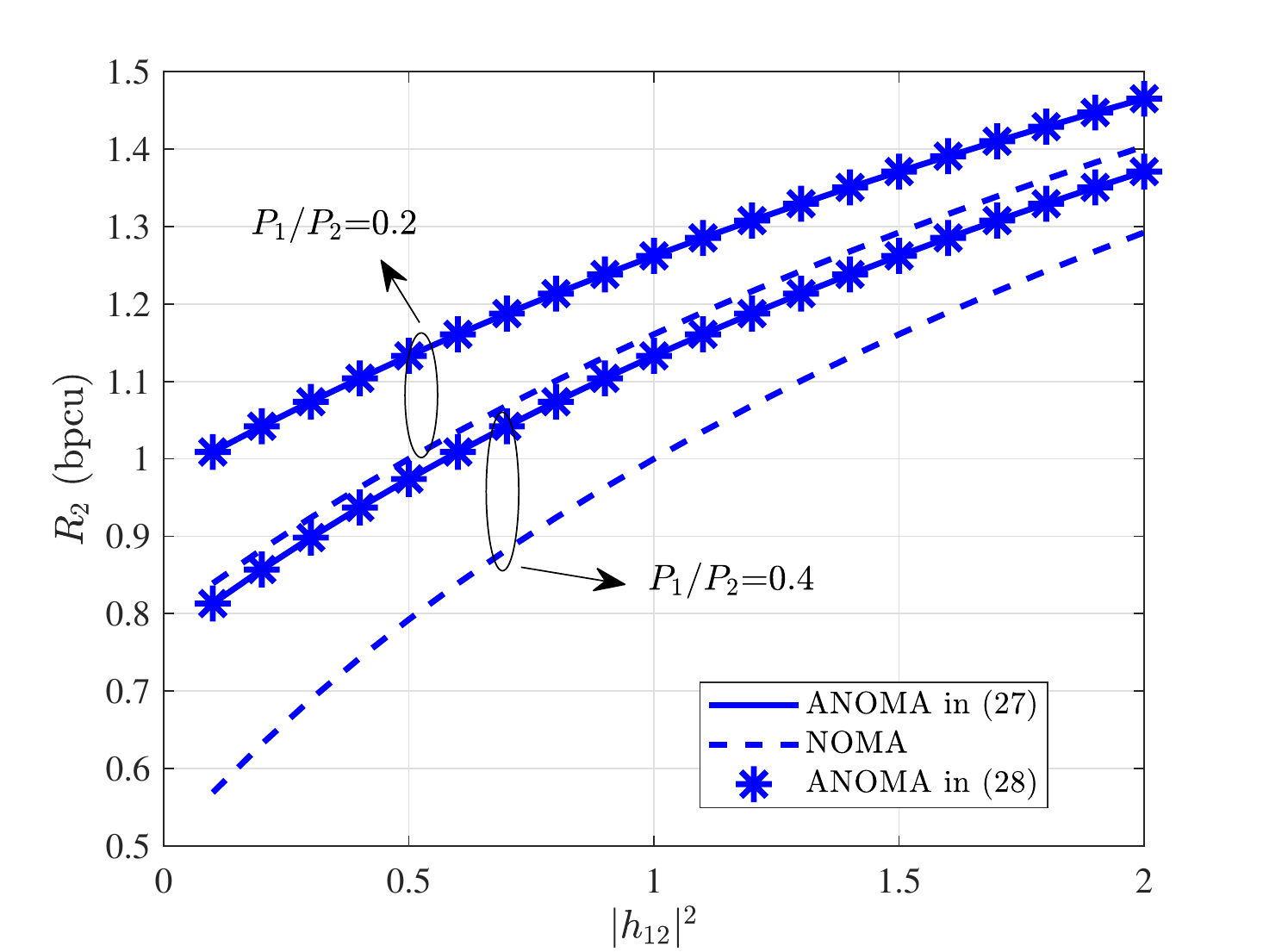}
	\caption{The throughput $R_{2}$ as a function of the channel gain $|h_{12}|^2$ for C-ANOMA and C-NOMA systems when $N = 10$, $\tau = 0.5$, $P_1 + P_2 = 5$, $P_r = 2$, $|h_2|^2 = 1$.}
	\label{fig_rate_vs_h_R2}
\end{figure}

\section{Numerical Results}\label{sec:numericalResults}
In this section, we present numerical results to compare the throughputs and power consumptions of C-NOMA and C-ANOMA systems.

First, we compare the throughputs of User 1 in the C-NOMA and C-ANOMA systems with different ratios of $P_1$ to $P_2$ in Fig. \ref{fig_rate_vs_h_R2-1}. The curves of ``ANOMA in \eqref{eqR2->1matrixform}/\eqref{eqR2->1analytical}/\eqref{eqR1matrixform}/\eqref{eqR1analytical}'' are derived directly from the expressions in \eqref{eqR2->1matrixform}/\eqref{eqR2->1analytical}/\eqref{eqR1matrixform}/\eqref{eqR1analytical}. In Fig.~\ref{fig_rate_vs_h_R2-1}, it is shown that the throughputs calculated by \eqref{eqR2->1analytical} and \eqref{eqR1analytical} completely align with the results of \eqref{eqR2->1matrixform} and \eqref{eqR1matrixform}, respectively, which verifies the correctness of \eqref{eqR2->1analytical} and \eqref{eqR1analytical}. Besides, it is demonstrated that the throughputs $R_{2\rightarrow 1}$ in the C-ANOMA systems are higher than those in the C-NOMA systems. Moreover, Fig. \ref{fig_rate_vs_h_R2-1} shows that the throughputs in both C-ANOMA and C-NOMA systems increase with the channel gain $|h_1|^2$. More specifically, the gaps of the throughput $R_{2\rightarrow 1}$ between the C-ANOMA and C-NOMA systems grow wider as $|h_1|^2$ increases. Fig. \ref{fig_rate_vs_h_R2-1} also shows that $R_1^{\mathrm{ANOMA}}$ is less than but very close to $R_1^{\mathrm{NOMA}}$ even for a relatively small frame length $N = 10$, especially when $|h_1|^2$ is small.

In Fig. \ref{fig_rate_vs_h_R2}, we compare the the throughput performance of User 2 in C-NOMA and C-ANOMA systems for different ratios of $P_1$ to $P_2$. Fig. \ref{fig_rate_vs_h_R2} verifies the correctness of Theorem \ref{theoremR2ANOMA} and the superiority of the C-ANOMA systems over the C-NOMA systems in terms of the throughput performance. One phenomenon we need to point out is that the throughput gaps between the C-ANOMA and C-NOMA systems shrink as $|h_{12}|^2$ increases, which is different from Fig. \ref{fig_rate_vs_h_R2-1}. Note that the throughput of User 2 depends on both the broadcast link from BS and the relay link from User 1. The sampling diversity can only be obtained through the asynchronous transmission from the broadcast link. As $|h_{12}|^2$ increases, the relay link becomes more and more dominant in calculating the throughput of User 2. Accordingly, the throughput gain from the sampling diversity becomes less and less noticeable as $|h_{12}|^2$ increases while the channel gain of the broadcast link $|h_2|^2$ is constant.

\begin{figure}[!ht b]
	\centering
	\includegraphics[width=4.5in]{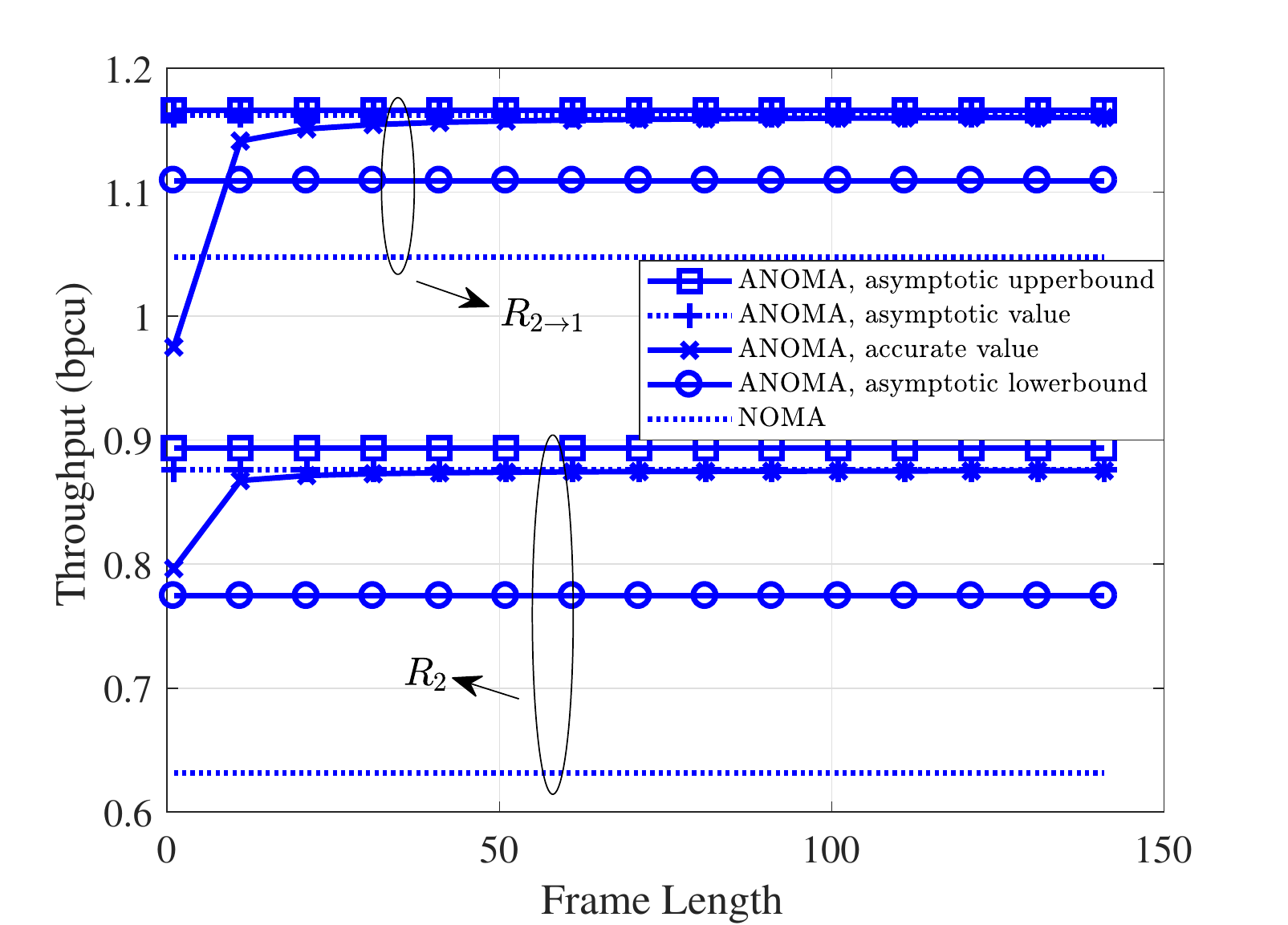}
	\caption{The throughputs $R_{2}$ and $R_{2\rightarrow 1}$ as functions of the frame length $N$ for C-ANOMA and C-NOMA systems when $\tau = 0.5$, $P_1 = 1.5$, $P_2 = 3.5$, $P_r = 2$, $|h_1|^2 = 1$, $|h_{2}|^2 = 0.8$, $|h_{12}|^2 = 1$.}
	\label{fig_R_vs_N}
\end{figure}

\begin{figure}[!htb]
	\centering
	\includegraphics[width=4.5in]{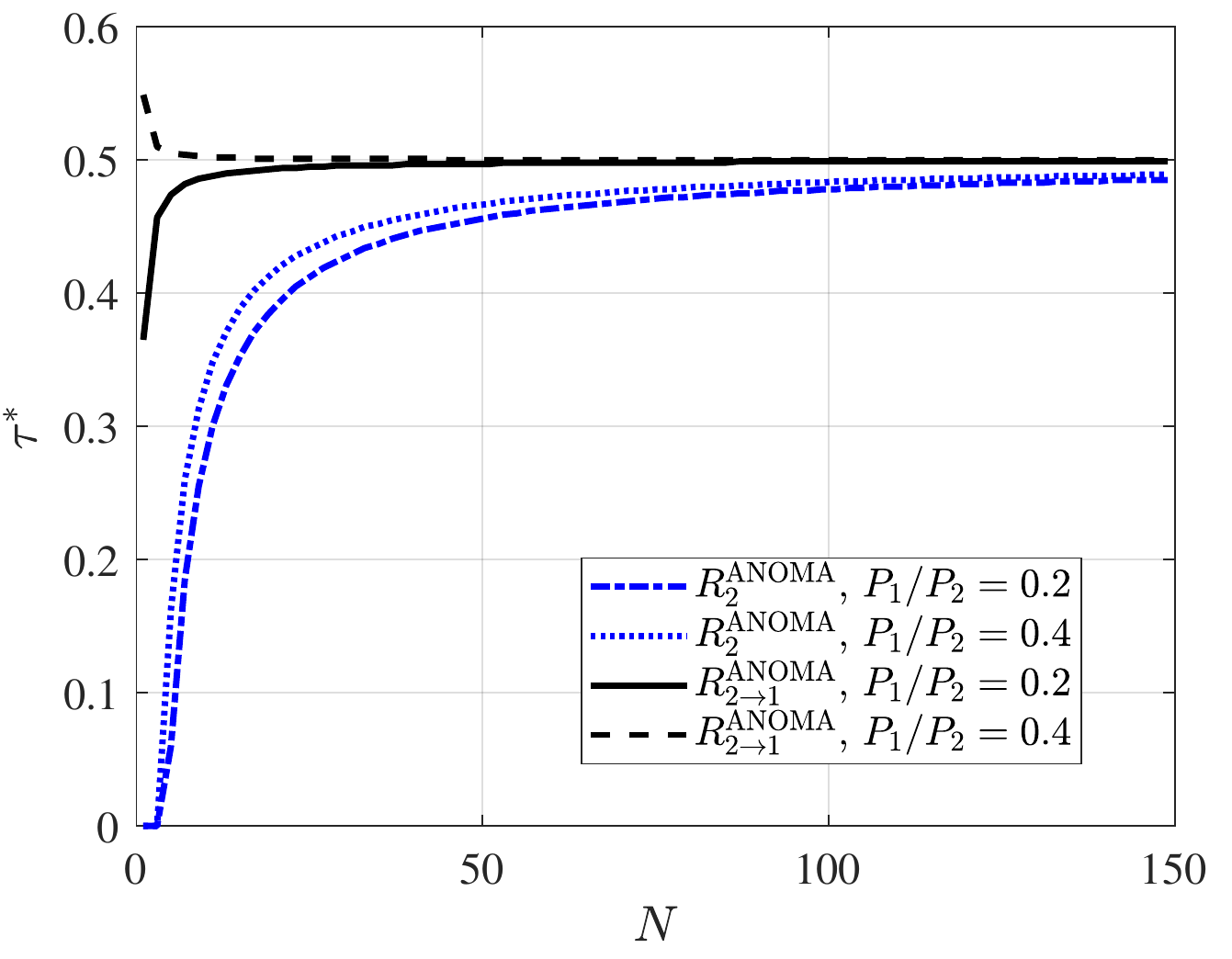}
	\caption{The optimal normalized timing mismatch $\tau^*$ to maximize the throughputs $R_2^{\mathrm{ANOMA}}$ and $R_{2\rightarrow 1}^{\mathrm{ANOMA}}$ as a function of the frame length $N$ when $P_1 + P_2 = 5$, $P_r = 2$, $|h_1|^2 = 1$, $|h_2|^2 = 0.5$, $|h_{12}|^2 = 2$.}
	\label{fig_optimal_tau}
\end{figure}

\begin{figure}[!htb]
	\centering
	\includegraphics[width=6in]{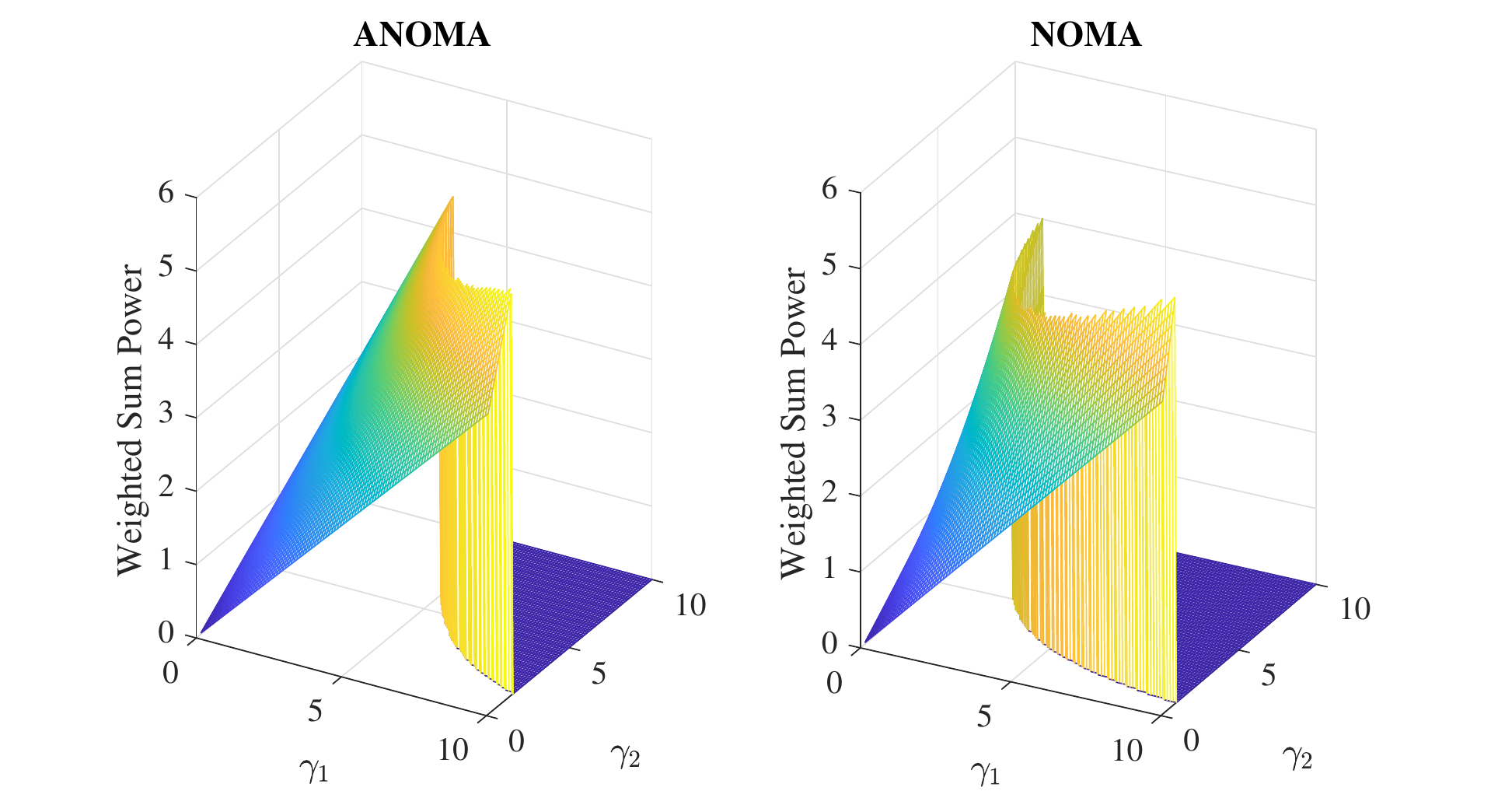}
	\caption{The minimized weighted sum power under the QoS constraints as a function of the target SINRs, $\gamma_1$ and $\gamma_2$, for the C-NOMA and C-ANOMA systems when $\tau = 0.5$, $P_{s,\max} = 20$, $P_{r,\max} = 5$, $\omega_s = 0.2$, $\omega_r = 0.8$, $|h_1|^2 = 1$, $|h_2|^2 = 0.5$, $|h_{12}|^2 = 2$, $N = 100$.}
	\label{fig_sumpower}
\end{figure}
\begin{figure}[!htb]
	\centering
	\includegraphics[width=4.5in]{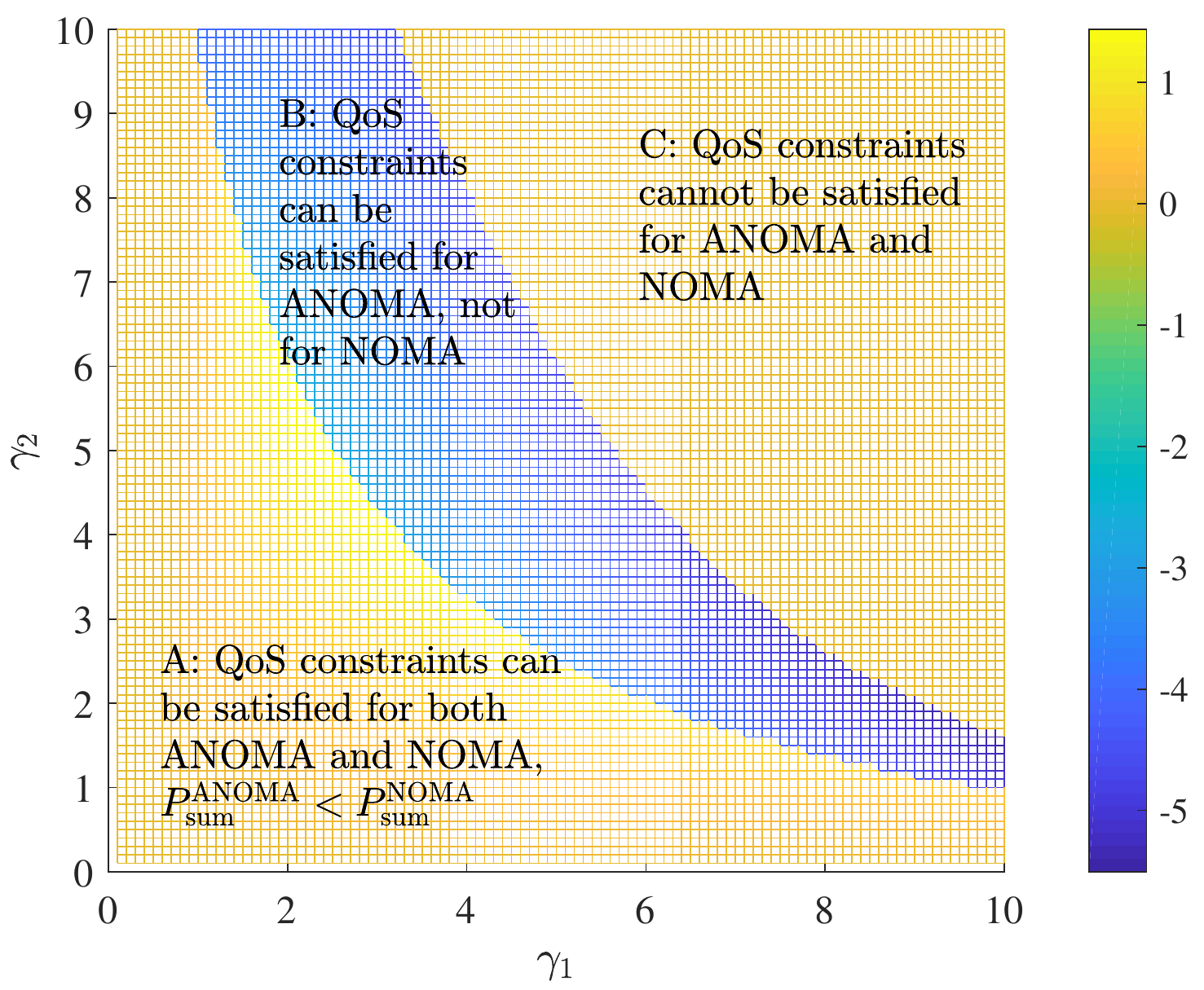}
	\caption{Illustration of the difference between the minimized weighted sum power in C-NOMA systems and that in C-ANOMA systems.}
	\label{fig_sumpower_dif}
\end{figure}

In Fig. \ref{fig_R_vs_N}, we show how the throughputs in C-ANOMA systems change with the frame length $N$. Since the expression for $R_1^{\mathrm{ANOMA}}$ in \eqref{eqR1analytical} is simple, the curves of $R_1^{\mathrm{ANOMA}}$ are omitted in Fig. \ref{fig_R_vs_N}. It is shown that as the frame length increases, the actual throughputs $R_{2\rightarrow 1}^{\mathrm{ANOMA}}$ and $R_2^{\mathrm{ANOMA}}$ converge to the asymptotic ones calculated by \eqref{eqR2->1asympANOMA} and \eqref{eqR2ANOMAasymp}, respectively. We note that the asymptotic throughputs, $R_{2\rightarrow 1,\mathrm{asymp}}^{\mathrm{ANOMA}}$ and $R_{2,\mathrm{asymp}}^{\mathrm{ANOMA}}$, perfectly approximate the actual throughputs, $R_{2\rightarrow 1}^{\mathrm{ANOMA}}$ and $R_2^{\mathrm{ANOMA}}$, when $N > 50$. And for both $R_{2\rightarrow 1}^{\mathrm{ANOMA}}$ and $R_2^{\mathrm{ANOMA}}$, the actual throughputs exceed their asymptotic lower bounds when $N > 20$. As a result, for $N > 20$, it is reasonable to use the lower bounds of the asymptotic throughputs as the constraints \eqref{suboptimal:eqR2->1>R2*} and \eqref{suboptimal:eqR2>R2*} in order to simplify the optimization problem. Besides, Fig. \ref{fig_R_vs_N} verifies Theorems \ref{ThmR2->1ANOMA>R2->1NOMA} and \ref{theoremR2ANOMA>R2NOMA} in addition to showing that the C-ANOMA systems outperform the C-NOMA systems for relatively small values of $N$. 

We also study the optimal design of C-ANOMA systems. Fig. \ref{fig_optimal_tau} shows the optimal normalized timing mismatch $\tau^*$ to maximize $R_2^{\mathrm{ANOMA}}$ or $R_{2\rightarrow 1}^{\mathrm{ANOMA}}$ as a function of the frame length $N$. In our simulation, $\tau^*$ is found by exhaustive search. Although $\tau^*$ varies a lot when $N$ is relatively small, $\tau^*$ converges to 0.5 steadily as $N$ increases for both $R_2^{\mathrm{ANOMA}}$ and $R_{2\rightarrow 1}^{\mathrm{ANOMA}}$ with different ratios of $P_1$ to $P_2$, as predicted by our analytical results. This is because the timing mismatch only exists in the asynchronous transmission in the broadcast phase and will affect $R_{2\rightarrow 1}^{\mathrm{ANOMA}}$ and $R_2^{\mathrm{ANOMA}}$ in the same way.

Moreover, we show the minimized weighted sum power under the QoS constraints as a function of target SINRs, $\gamma_1$ and $\gamma_2$, for C-NOMA and C-ANOMA systems in Fig. \ref{fig_sumpower}. We set $\omega_s$ and $\omega_r$ as 0.2 and 0.8, respectively, because the power consumption of the relay user with limited battery capacity has a higher priority in the power minimization problem. In Fig. \ref{fig_sumpower}, the weighted sum power is calculated by solving the power optimization problem \eqref{eq:suboptimal2} for the C-NOMA (setting $\tau = 0$) and C-ANOMA (setting $\tau = 0.5$) systems. In our simulation, we assume that BS and the relay user will stop transmission (i.e., $P_1 = P_2 = P_r = 0$) if the QoS constraints cannot be satisfied. For both C-NOMA and C-ANOMA systems, it is shown in Fig. \ref{fig_sumpower} that the weighted sum power increases with the target SINRs until BS and the relay user reach their power limits and stop transmission. To further compare the power consumptions, we calculate the difference of the weighted sum powers between the C-NOMA and C-ANOMA systems and provide the results in Fig. \ref{fig_sumpower_dif}. As shown in Fig.~\ref{fig_sumpower_dif}, C-ANOMA systems can consume less power compared with C-NOMA systems to guarantee the same QoS in the area A. In the area B, it is shown that C-ANOMA systems can still satisfy the QoS with limited transmit powers while C-NOMA systems cannot. When both $\gamma_1$ and $\gamma_2$ are large, i.e., the area C in Fig. \ref{fig_sumpower_dif}, neither C-NOMA nor ANOMA systems can satisfy the QoS with the limited transmit powers. 

\begin{figure}[!htb]
	\centering
	\includegraphics[width=4.5in]{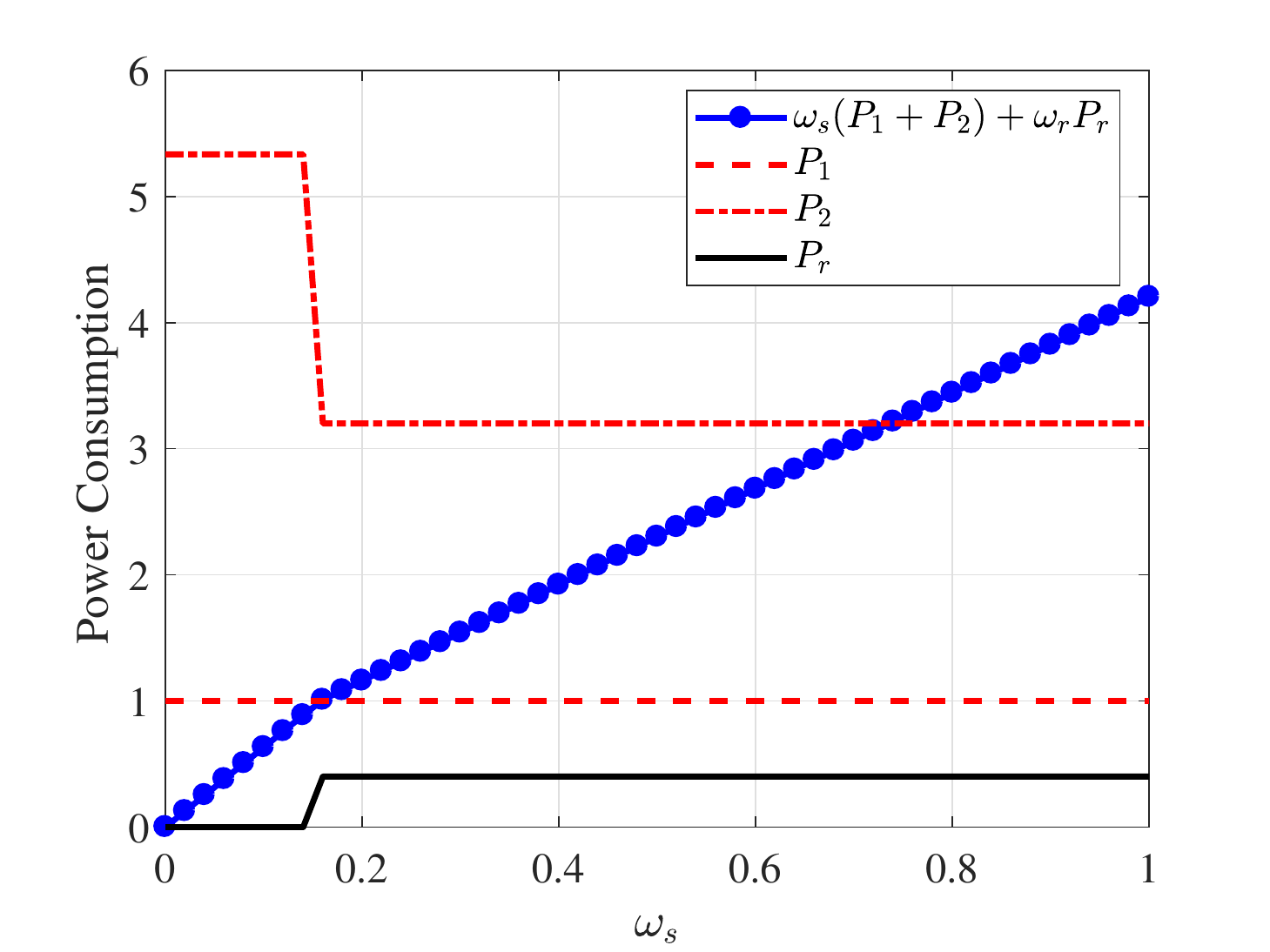}
	\caption{The power consumptions as functions of the weight allocated to the transmit power of BS, i.e., $\omega_s$, for the C-ANOMA systems when $\tau = 0.5$, $P_{s,\max} = 20$, $P_{r,\max} = 5$, $|h_1|^2 = 1$, $|h_2|^2 = 0.5$, $|h_{12}|^2 = 2$, $N = 100$.}
	\label{fig_power_vs_omega}
\end{figure}

Finally, we show how the power consumptions change with the weight $\omega_s$ in the power minimization problem in Fig. \ref{fig_power_vs_omega}. In our simulation, we set $\omega_r = 1 - \omega_s$. In Fig. \ref{fig_power_vs_omega}, the power allocated to User 1 does not change with $\omega_s$ as long as BS has enough transmit power to support the QoS of User 1. If $\omega_s$ is large, BS can save a large amount of power (decreases by about 2) under the help of the relay user (transmit power increases by about 0.5) because the channel of the relay link is better than that of the broadcast link between BS and User 2. When $\omega_s$ is small ($\omega_r$ is large), the relay user keeps silent to reduce energy consumption. When $\omega_s$ is large ($\omega_r$ is small), BS communicates with User~2 under the help of the relay user, which takes advantage of the relay link to complement the large path loss between BS and User~2. Hence, Fig. \ref{fig_power_vs_omega} shows that one can make a trade-off between the power consumption of BS and that of the relay user by adjusting the weight $\omega_s$. 

\section{Conclusion}\label{sectionConclusion}
In this paper, we study the half-duplex C-ANOMA systems with user relaying. We analytically prove that for a sufficiently large frame length, the strong user in C-ANOMA systems can achieve the same throughput as that in C-NOMA systems while the weak user in C-ANOMA systems benefits from the symbol-asynchronous transmission. Moreover, we analyze the optimal design of the C-ANOMA systems. As the frame length increases, the optimal timing mismatch converges to half of the symbol interval. Besides, we solve a weighted sum power minimization problem under QoS constraints. Numerical results demonstrate that C-ANOMA systems can consume less power to satisfy the same QoS requirements compared with C-NOMA systems.

\appendices
\section{Derivation of \eqref{eqR2->1analytical} and \eqref{eqR1analytical}}\label{prooflemma(1+mu_1)N}
Substituting $\mathbf{G}_1$ and $\mathbf{R}$ by their expressions, the matrix determinant term in \eqref{eqR1matrixform} becomes
	\begin{align}\label{eqLemma(1+mu_1)N}
	&\det\left(\mathbf{I}_{2N} + P_1|h_1|^2\mathbf{G}_1\mathbf{G}_1^H\mathbf{R}\right)\notag\\
	&= \det\left[\begin{smallmatrix}
	1+ P_1|h_1|^2 & P_1|h_1|^2(1-\tau) &0 &\cdots &\cdots &0\\
	0 &1 &0 &0 &\cdots &0\\
	0 & P_1|h_1|^2\tau &1+ P_1|h_1|^2 & P_1|h_1|^2(1-\tau) &\cdots &0\\
	\vdots &\ddots &\ddots &\ddots &\ddots &\vdots\\
	0 &\cdots &0 &1 &0 &0\\
	0 &\cdots &0 & P_1|h_1|^2\tau &1+ P_1|h_1|^2 &P_1|h_1|^2(1-\tau)\\
	0 &\cdots &\cdots &0 &0 &1
	\end{smallmatrix}\right]_{2N \times 2N}\notag\\
	&\stackrel{(a)}{=} \det\left[\begin{smallmatrix}
	1+P_1|h_1|^2 &P_1|h_1|^2(1-\tau) &0 &\cdots &\cdots &0\\
	0 &1 &0 &0 &\cdots &0\\
	0 &P_1|h_1|^2\tau &1+P_1|h_1|^2 &P_1|h_1|^2(1-\tau) &\cdots &0\\
	\vdots &\ddots &\ddots &\ddots &\ddots &\vdots\\
	0 &\cdots &0 &0 &1 &0\\
	0 &\cdots &0 &0 &P_1|h_1|^2\tau &1+P_1|h_1|^2\\
	\end{smallmatrix}\right]_{(2N-1) \times (2N-1)}\notag\\
	&\stackrel{(b)}{=} \left(1\!+\!P_1|h_1|^2\right) \det\left[\begin{smallmatrix}
	1+P_1|h_1|^2 &P_1|h_1|^2(1-\tau) &0 &\cdots &\cdots &0\\
	0 &1 &0 &0 &\cdots &0\\
	0 &P_1|h_1|^2\tau &1+P_1|h_1|^2 &P_1|h_1|^2(1-\tau) &\cdots &0\\
	\vdots &\ddots &\ddots &\ddots &\ddots &\vdots\\
	0 &\cdots &0 &P_1|h_1|^2\tau &1+P_1|h_1|^2 &P_1|h_1|^2(1-\tau)\\
	0 &\cdots &\cdots &0 &0 &1
	\end{smallmatrix}\right]_{(2N-2) \times (2N-2)}\notag\\
	&=\cdots\stackrel{(c)}{=} (1+P_1|h_1|^2)^N,
	\end{align}
where $(a)$ and $(b)$ are derived by applying the cofactor expansion~\cite{poole2014linear}, $(c)$ is derived by applying the cofactor expansion iteratively. Thus, Eq. \eqref{eqR1analytical} is obtained.

According to \eqref{eqR2->1matrixform}, we have
\begin{align}\label{eqR2->1matrixformAppendix}
	&R_{2\rightarrow 1}^{\mathrm{ANOMA}} \notag\\
    &= \frac{1}{2N+\tau}\log\det\left[\mathbf{I}_{2N} + \left(\mathbf{I}_{2N} + P_1|h_1|^2\mathbf{G}_1\mathbf{G}_1^H\mathbf{R}\right)^{-1} P_2|h_1|^2\mathbf{G}_2\mathbf{G}_2^H\mathbf{R}\right]\notag\\
	&= \frac{1}{2N+\tau}\log\det\left[\left(\mathbf{I}_{2N} +  P_1|h_1|^2\mathbf{G}_1\mathbf{G}_1^H\mathbf{R} \right)^{-1}\left(\mathbf{I}_{2N} +  P_1|h_1|^2\mathbf{G}_1\mathbf{G}_1^H\mathbf{R} +  P_2|h_1|^2\mathbf{G}_2\mathbf{G}_2^H\mathbf{R}\right)\right]\notag\\
	&= \frac{1}{2N+\tau}\log\det\left(\mathbf{I}_{2N} + P_1|h_1|^2\mathbf{G}_1\mathbf{G}_1^H\mathbf{R} + P_2|h_1|^2\mathbf{G}_2\mathbf{G}_2^H\mathbf{R}\right)\notag\\
	&\ \ \ - \frac{1}{2N+\tau}\log\det\left(\mathbf{I}_{2N} + P_1|h_1|^2\mathbf{G}_1\mathbf{G}_1^H\mathbf{R} \right)\notag\\
	&\stackrel{(a)}{=} \frac{1}{2N+\tau}\log\det\left(\mathbf{I}_{2N} + \mathbf{H}\mathbf{R}\right) - \frac{1}{2N+\tau}\log\det\left(\mathbf{I}_{2N} + P_1|h_1|^2\mathbf{G}_1\mathbf{G}_1^H\mathbf{R} \right),
\end{align}
where $(a)$ is derived because $\mathbf{G}_i\mathbf{G}_i^H$ is a $2N$-by-$2N$ matrix whose odd (if $i = 1$) or even (if $i = 2$) diagonal elements are 1 and all the others are 0, and $\mathbf{H} = |h_1|^2\cdot\mathrm{diag}\left(\left[P_1, P_2, \cdots, P_1, P_2\right]\right)$.

According to Theorem 1 in \cite{zou2018analysis}, the term $\log\det\left(\mathbf{I}_{2N} + \mathbf{HR}\right)$ in \eqref{eqR2->1matrixformAppendix} can be written as
\begin{align}\label{eqAppendixAciteformprevpaper}
\log\det\left(\mathbf{I}_{2N} + \mathbf{HR}\right) &= N\log\left(\mu_1\mu_2\right) + \log \frac{\left(r_1^{N+1} - r_2^{N+1}\right) + \tau^2\left(r_1^N - r_2^N\right)}{r_1 - r_2},
\end{align}
where 
\begin{align}
\mu_1 &= P_1|h_1|^2, \mu_2 = P_2|h_1|^2,Q=2\tau(1-\tau),\\
r_1 &=
\frac{\mu_1^{-1}\! +\! \mu_2^{-1}\! +\! \mu_1^{-1}\mu_2^{-1}\! +\! Q+ \sqrt{\left[\mu_1^{-1} + \mu_2^{-1} + \mu_1^{-1}\mu_2^{-1} + Q\right]^2\! -\! Q^2}}{2},\\
r_2 &=
\frac{\mu_1^{-1}\! +\! \mu_2^{-1}\! +\! \mu_1^{-1}\mu_2^{-1}\! +\! Q- \sqrt{\left[\mu_1^{-1} + \mu_2^{-1} + \mu_1^{-1}\mu_2^{-1} + Q\right]^2\! -\! Q^2}}{2}.
\end{align}

Thus, Eq. \eqref{eqR2->1analytical} can be easily derived according to \eqref{eqLemma(1+mu_1)N} and \eqref{eqAppendixAciteformprevpaper}.

\section{Proof of Theorem \ref{ThmR2->1ANOMA>R2->1NOMA}}\label{proofThmR2ANOMA>R2NOMA}
\begin{IEEEproof}
	According to Corollary 1 in~\cite{zou2018analysis}, we have
	\begin{align}\label{eqReferFromPrevPaper}
	    \lim_{N\rightarrow \infty} \frac{1}{N+\tau}\log \frac{\left(r_1^{N+1} - r_2^{N+1}\right) + \tau^2\left(r_1^N - r_2^N\right)}{r_1 - r_2} = \log r_1.
	\end{align}
	As a result, the throughput of User~2 for $N\rightarrow \infty$ is calculated as
	\begin{align}
	\nonumber
	R_{2,\mathrm{asymp}}^{\mathrm{ANOMA}} &= \frac{1}{2}\log\left(\frac{\mu_1\mu_2 r_1}{1 + \mu_1}\right)\notag\\
	\nonumber
	&= \frac{1}{2}\log \left(\frac{1 + \mu_1 + \mu_2 + \mu_1\mu_2Q+\sqrt{\left(1+\mu_1+\mu_2\right)^2 + 2 \left(1+\mu_1 +\mu_2\right)\mu_1\mu_2Q}}{2(1+\mu_1)} \right),\label{eqproofR2->1asympANOMA}
	\end{align}
	\normalsize
	where $\mu_1 > 0$, $\mu_2 > 0$, $\tau \in [0, 1)$, and $Q = 2\tau(1-\tau) > 0$. One can easily derive
	\begin{align*}
	1 + \mu_1 + \mu_2 &\le \sqrt{\left(1+\mu_1+\mu_2\right)^2 + 2 \left(1+\mu_1 +\mu_2\right)\mu_1\mu_2Q}\\
	& = \sqrt{\left(1+\mu_1+\mu_2 + \mu_1\mu_2Q\right)^2 - \left(\mu_1\mu_2Q\right)^2} \le 1+\mu_1+\mu_2+\mu_1\mu_2Q,
	\end{align*}
	\normalsize
	and the equal sign is achieved if and only if $\tau = 0$.
	
	As a result,
	\begin{align}
	\nonumber
	\frac{1}{2}\log\left(\frac{1 + \mu_1 + \mu_2 + 0.5\mu_1\mu_2Q}{1+\mu_1}\right)\le R_{2,\mathrm{asymp}}^{\mathrm{ANOMA}} \le \frac{1}{2}\log\left(\frac{1 + \mu_1 + \mu_2 + \mu_1\mu_2Q}{1+\mu_1}\right).
	\end{align}
	
	Note that,
	\begin{align}
	\frac{1}{2}\log\left(\frac{1 + \mu_1 + \mu_2 + 0.5\mu_1\mu_2Q}{1+\mu_1}\right)\ge \frac{1}{2}\log\left(\frac{1 + \mu_1 + \mu_2}{1+\mu_1}\right) = R_2^{\mathrm{NOMA}},
	\end{align} 
	where the equal sign is achieved if and only if $\tau=0$. The proof is complete.
\end{IEEEproof}

\section{Proof of Theorem \ref{theoremR2ANOMA}}\label{prooftheoremR2ANOMA}
\begin{IEEEproof}
According to \eqref{eqR2ANOMAmatrix}, the throughput of User 2 is given by
\begin{align}
&R_2^{\mathrm{ANOMA}} \notag\\
&= \frac{1}{2N+\tau}\log\det\left[\mathbf{I}_{3N} + \left(\mathbf{R_N} + \mathbf{W}_1\mathbf{W}_1^H\right)^{-1}\mathbf{W}_2\mathbf{W}_2^H\right]\notag\\
&= \frac{1}{2N+\tau}\log\det\left(\mathbf{I}_{3N} + \left[\begin{smallmatrix}
\mathbf{R} + P_1|h_2|^2\mathbf{RG}_1\mathbf{G}_1^H\mathbf{R}^H &\mathbf{0}\\\mathbf{0} &\mathbf{I}_N
\end{smallmatrix}\right]^{-1}\left[\begin{smallmatrix}
P_2|h_2|^2\mathbf{RG}_2\mathbf{G}_2^H\mathbf{R}^H &\sqrt{P_2P_r}h_2\bar{h}_{12}\mathbf{RG}_2\\
\sqrt{P_2P_r}h_{12}\bar{h}_2\mathbf{G}_2^H\mathbf{R}^H &P_r|h_{12}|^2\mathbf{I}_N
\end{smallmatrix}\right]\right)\notag\\
&= \frac{1}{2N+\tau}\log\det\left[\begin{smallmatrix}
\mathbf{I}_{2N} + P_2|h_2|^2\left(\mathbf{R} + P_1|h_2|^2\mathbf{RG}_1\mathbf{G}_1^H\mathbf{R}\right)^{\!-1\!}\mathbf{RG}_2\mathbf{G}_2^H\mathbf{R}\ &\sqrt{P_2P_r}h_2\bar{h}_{12}\left(\mathbf{R} + P_1|h_2|^2\mathbf{RG}_1\mathbf{G}_1^H\mathbf{R}\right)^{\!-1\!}\mathbf{RG}_2\\
\sqrt{P_2P_r}h_{12}\bar{h}_2\mathbf{G}_2^H\mathbf{R} &\mathbf{I}_{N}+P_r|h_{12}|^2\mathbf{I}_N
\end{smallmatrix}\right]\notag\\
&\stackrel{(a)}{=} \frac{1}{2N+\tau}\log\left\{\det\left[(1+P_r|h_{12}|^2)\mathbf{I}_N\right]\det\left[\mathbf{I}_{2N} + P_2|h_2|^2\left(\mathbf{I}_{2N} + P_1|h_2|^2\mathbf{G}_1\mathbf{G}_1^H\mathbf{R}\right)^{-1}\mathbf{G}_2\mathbf{G}_2^H\mathbf{R}\right.\right.\notag\\
&\ \ \ \left.\left. - \frac{P_2P_r|h_2|^2|h_{12}|^2}{1+P_r|h_{12}|^2}\left(\mathbf{I}_{2N} + P_1|h_2|^2\mathbf{G}_1\mathbf{G}_1^H\mathbf{R}\right)^{-1}\mathbf{G}_2\mathbf{G}_2^H\mathbf{R} \right]\right\}\notag\\
&= \frac{N}{2N+\tau}\log\left(1+P_r|h_{12}|^2\right) + \frac{1}{2N+\tau}\log\det\left\{\left(\mathbf{I}_{2N} + P_1|h_2|^2\mathbf{G}_1\mathbf{G}_1^H\mathbf{R}\right)^{-1}\right.\notag\\
&\ \ \ \cdot\left.\left[\mathbf{I}_{2N} + P_1|h_2|^2\mathbf{G}_1\mathbf{G}_1^H\mathbf{R} + \left(P_2|h_2|^2- \frac{P_2P_r|h_2|^2|h_{12}|^2}{1+P_r|h_{12}|^2}\right)\left(\mathbf{I}_{2N} + P_1|h_2|^2\mathbf{G}_2\mathbf{G}_2^H\mathbf{R}\right)\right]\right\}\notag\\
&= \frac{N}{2N+\tau}\log\left(1+P_r|h_{12}|^2\right) + \frac{1}{2N+\tau}\log\det\left(\mathbf{I}_{2N} + P_1|h_2|^2\mathbf{G}_1\mathbf{G}_1^H\mathbf{R}\right)^{-1}\notag\\
&\ \ \ + \frac{1}{2N+\tau}\log\det\left(\mathbf{I}_{2N} + P_1|h_2|^2\mathbf{G}_1\mathbf{G}_1^H\mathbf{R} + \frac{P_2|h_2|^2}{1+P_r|h_{12}|^2}\mathbf{G}_2\mathbf{G}_2^H\mathbf{R}\right)\notag\\
&= \frac{N}{2N\!+\!\tau}\log\left(1\!+\!P_r|h_{12}|^2\right) - \frac{\log\det\left(\mathbf{I}_{2N} + P_1|h_2|^2\mathbf{G}_1\mathbf{G}_1^H\mathbf{R}\right)}{2N\!+\!\tau}\!+\! \frac{\log\det(\mathbf{I}_{2N} \!+\! \tilde{\mathbf{H}}\mathbf{R})}{2N\!+\!\tau},\label{eqproofR2ANOMAmid}
\end{align}
where $(a)$ is derived by applying the determinant of the block matrix, i.e., if $\mathbf{D}$ is invertible,
\begin{align}
\det\left(\begin{smallmatrix}
\mathbf{A} &\mathbf{B}\\ \mathbf{C} &\mathbf{D}
\end{smallmatrix}\right) = \det(\mathbf{D})\det(\mathbf{A-BD}^{-1}\mathbf{C})
\end{align} 
and $\tilde{\mathbf{H}} = \mathrm{diag}\left(\left[P_1|h_2|^2, \frac{P_2|h_2|^2}{1+P_r|h_{12}|^2}, \cdots P_1|h_2|^2, \frac{P_2|h_2|^2}{1+P_r|h_{12}|^2}\right]\right)$.

Applying \eqref{eqLemma(1+mu_1)N} and \eqref{eqAppendixAciteformprevpaper}, Eq. \eqref{eqproofR2ANOMAmid} can be rewritten as \eqref{eqR2ANOMA}. The proof is complete.
\end{IEEEproof}

\section{Proof of Theorem~\ref{theoremR2ANOMA>R2NOMA}}\label{proofTheoremR2ANOMA>R2NOMA}
\begin{IEEEproof}
Applying \eqref{eqReferFromPrevPaper}, the throughput of User~2 for $N\rightarrow\infty$ is computed as
\begin{align}
    R_{2,\mathrm{asymp}}^{\mathrm{ANOMA}} =& \frac{1}{2}\log\left(\frac{P_1P_2|h_{2}|^4}{1 + P_1|h_2|^2}z_1\right)\notag\\
    \stackrel{(a)}{=}& \frac{1}{2}\log\!\left[\frac{1 + P_r|h_{12}|^2}{2} + \frac{P_2|h_2|^2 + P_1P_2|h_2|^4Q}{2(1+ P_1|h_2|^2)}\right.\notag\\
	&+ \! \left. \!\frac{1}{2}\sqrt{\!\left(1 + P_r|h_{12}|^2 + \frac{ P_2|h_2|^2 \!+\! P_1P_2|h_2|^4Q }{1+ P_1|h_2|^2}\right)^2\!-\! \left(\frac{P_1P_2|h_2|^4Q}{1+ P_1|h_2|^2}\right)^2}\right],
\end{align}
where $(a)$ is derived by replacing $z_1$ with its expression in \eqref{eqz1expression}. Since $Q \ge 0$,
\begin{align*}
    1 + P_r|h_{12}|^2 &+ \frac{ P_2|h_2|^2 \!+\! P_1P_2|h_2|^4Q }{1+ P_1|h_2|^2} \\
    &\ge \sqrt{\!\left(1 + P_r|h_{12}|^2 + \frac{ P_2|h_2|^2 \!+\! P_1P_2|h_2|^4Q }{1+ P_1|h_2|^2}\right)^2\!-\! \left(\frac{P_1P_2|h_2|^4Q}{1+ P_1|h_2|^2}\right)^2}\\
    &= \sqrt{\!\left(1 + P_r|h_{12}|^2 \!+\! \frac{ P_2|h_2|^2}{1+ P_1|h_2|^2}\right)^2 + \frac{2P_1P_2|h_2|^4Q }{1+ P_1|h_2|^2}\left(1 + P_r|h_{12}|^2 + \frac{ P_2|h_2|^2}{1+ P_1|h_2|^2}\right)}\\
    &\ge 1 + P_r|h_{12}|^2 \!+\! \frac{ P_2|h_2|^2}{1+ P_1|h_2|^2},
\end{align*}
where the equal signs are achieved if and only if $Q = 0$ which results in $\tau=0$. The proof is complete.
\end{IEEEproof}

%
%
%


\ifCLASSOPTIONcaptionsoff
  \newpage
\fi



{\small
\bibliographystyle{IEEEtran}
\bibliography{IEEEabrv,IEEEexample}}
\end{document}